\documentclass[12pt]{article}
\usepackage{graphicx}

\makeatletter
\newcommand\eqsecnum{
\@newctr{equation}[section]

\renewcommand\theequation{\arabic{section}.\arabic{equation}}%
}
\newcommand{\vereq}[2]{\lower3pt\vbox{\baselineskip1.5pt \lineskip1.5pt
\ialign{$\m@th#1\hfill##\hfil$\crcr#2\crcr\sim\crcr}}}

\renewcommand\appendix{\par
  \setcounter{section}{0}%
  \setcounter{subsection}{0}%
  \gdef\thesection{\@Alph\c@section}%
\renewcommand\theequation{\Alph{section}.\arabic{equation}}
}

\newcount\@indentflag \global\@indentflag=1 %
\newdimen\@eqtoeqnum \@eqtoeqnum=6pt %
\def\@indentamount{%
\ifcase\@indentflag 0pt\or\@centering\or0pt plus1fil\fi\relax
}
\def\FL{\global\@indentflag=0 }
\def\FR{\global\@indentflag=2 }
\def\@eqnnum{\hbox{\reset@font\rm(\theequation)}}
\let\make@eqnnum=\@eqnnum %
\def\eqnum#1{\dec@eqnnum \global\def\make@eqnnum{\reset@font\rm(#1)}%
\def\@currentlabel{#1}%
}
\def\inc@eqnnum{\addtocounter{equation}{1}}
\def\dec@eqnnum{\addtocounter{equation}{-1}}
\newbox\@testboxa
\newbox\@testboxb
\def\equation{\par\vskip-\lastskip\vskip\abovedisplayskip
\inc@eqnnum\let\@currentlabel=\theequation
\setbox\@testboxa=\hbox\bgroup\hskip\@totalleftmargin\hskip\@indentamount
\hbox\bgroup$\displaystyle
}
\def\endequation{$\egroup\hskip\@centering\egroup %
\setbox\@testboxb=\hbox{\make@eqnnum}%
\bgroup
\@tempdima\wd\@testboxa \advance\@tempdima by\wd\@testboxb
\ifcase\@indentflag
\advance\@tempdima by\@eqtoeqnum
\ifdim\@tempdima<\hsize %
\def\@tempa{0}%
\else
\def\@tempa{1}%
\fi
\or
\advance\@tempdima by2\@eqtoeqnum
\ifdim\@tempdima<\hsize %
\def\@tempa{0}%
\else %
\@tempdima\wd\@testboxa \advance\@tempdima by\wd\@testboxb
\advance\@tempdima by\@eqtoeqnum
\ifdim\@tempdima<\hsize %
\def\@tempa{0}%
\setbox\@testboxa\hbox{\hfill\box\@testboxa\kern\@eqtoeqnum}%
\else
\def\@tempa{1}%
\fi
\fi
\or
\advance\@tempdima by2\@eqtoeqnum
\ifdim\@tempdima<\hsize %
\def\@tempa{0}%
\setbox\@testboxb=\hbox{\kern\@eqtoeqnum\make@eqnnum}%
\else
\def\@tempa{1}%
\fi
\fi
\ifnum\@tempa=0 %
\hbox to\hsize{\unhbox\@testboxa\box\@testboxb}%
\else %
\vbox{\hbox to\hsize{\unhbox\@testboxa}%
\vskip6pt %
\hbox to\hsize{\hfil\box\@testboxb}}%
\fi
\egroup
\global\let\make@eqnnum\@eqnnum %
\vskip\belowdisplayskip\noindent\global\@indentflag=1 \global\@ignoretrue
}
\def\eqnarray{\par\vskip-\lastskip\vskip\abovedisplayskip
\inc@eqnnum\let\@currentlabel=\theequation
\global\@eqnswtrue\m@th
\global\@eqcnt\z@
\tabskip\@totalleftmargin\advance\tabskip by\@indentamount\let\\\@eqncr
\halign to\hsize\bgroup\hskip\@centering
$\displaystyle\tabskip\z@{##{}}$&\global\@eqcnt\@ne
\hfil${{}##{}}$\hfil
&\global\@eqcnt\tw@ $\displaystyle\tabskip\z@{##}$\hfil
\tabskip\@centering \if@eqnsw\phantom{\make@eqnnum\kern\@eqtoeqnum}\fi
&\llap{##}\tabskip\z@\cr}
\def\endeqnarray{%
\@@eqncr\egroup
\vskip\belowdisplayskip\noindent
\dec@eqnnum\global\@indentflag=1
\global\let\make@eqnnum\@eqnnum %
\global\@ignoretrue
}
\newbox\tempboxa
\newdimen\captionboxsubcount
\def\capsize#1{\captionboxsubcount=#1pt}
\newdimen\captionboxsub
\captionboxsub=\hsize \advance\captionboxsub by -\captionboxsubcount
\advance\captionboxsub by -\captionboxsubcount
\long\def\@makecaption#1#2{
 \setbox\@tempboxa\hbox{\footnotesize #1: #2}
 \ifdim \wd\@tempboxa >\captionboxsub
\rightskip=\captionboxsubcount \leftskip=\captionboxsubcount
  \footnotesize #1: #2
\else \hbox to\hsize{\hfil\box\@tempboxa\hfil}
 \fi}

\makeatother
\capsize{30}

\eqsecnum

\def\beas{\begin{eqnarray*}}
\def\eeas{\end{eqnarray*}}
\def\be{\begin{eqnarray}}
\def\ee{\end{eqnarray}}
\def\bq{\begin{equation}}
\def\eq{\end{equation}}
\def\ben{\begin{enumerate}}\def\een{\end{enumerate}}

\def\del{\partial}

\def\gp{\gamma_{||}}\def\gn{\gamma_\perp}

\def\m{\mu}

\def\g{\gamma}

\def\p{\psi}
\def\tr{{\rm tr}\, }

\def\g{\gamma}

\def\m{\mu}

\def\parallel{{| \hskip-0.03cm |}}

\def\La{{\cal L}}

\def\ln{{\rm ln}}

\def\A0{A_0}

\def\roughly#1{\mathrel{\raise.3ex\hbox{$#1$\kern-.75em%
\lower1ex\hbox{$\sim$}}}}

\renewcommand{\thefootnote}{\fnsymbol{footnote}}
\renewcommand{\thefootnote}{\arabic{footnote}}
\def\plb #1 {Phys.~Lett.~B~{\bf #1}\, }
\def\np #1 {Nucl.~Phys.~{\bf #1}\, }
\def\prd #1 {Phys.~Rev.~D~{\bf #1}\, }
\def\prc #1 {Phys.~Rev.~C~{\bf #1}\, }
\def\prl #1 {Phys.~Rev.~Lett.~{\bf #1}\, }

\begin{document}

\renewcommand{\thefootnote}{\arabic{footnote}}
\setcounter{footnote}{0}

\begin{flushright}
\begin{minipage}{3cm}
\begin{flushleft}
DPNU-04-17 \\
SNUTP-04-019
\\
\today
\end{flushleft}
\end{minipage}
\end{flushright}

\begin{center}
{\LARGE\bf
An Effective Field Theory \\
at Finite Density}

\date{\today}

\vskip 1cm
Masayasu Harada$^{\rm(a)}$\footnote{%
  E-mail:harada@eken.phys.nagoya-u.ac.jp},
Dong-Pil Min$^{\rm(b)}$\footnote{E-mail:dpmin@phya.snu.ac.kr},
Tae-Sun Park$^{\rm(c)}$\footnote{E-mail:tspark@kias.re.kr},
Chihiro Sasaki$^{\rm(a)}$\footnote{%
  E-mail:sasaki@eken.phys.nagoya-u.ac.jp}
and
Chaejun Song$^{\rm(d)}$\footnote{%
  E-mail:chaejun@charm.physics.pusan.ac.kr}
\end{center}

\vskip 0.5cm

\begin{center}
$^{\rm(a)}${\it Department of Physics, Nagoya University,
  Nagoya 464-8602, Japan}\\
$^{\rm(b)}${\it School of Physics,
  Seoul National University, 151-747, Korea}\\
$^{\rm(c)}${\it School of Physics, Korea Institute for Advanced Study,
  Seoul 130-012, Korea}\\
$^{\rm(d)}${\it Department of Physics, Pusan National University,
  Busan 609-735, Korea
}
\end{center}

\vskip 0.5cm

\begin{abstract}

An effective theory to treat the dense
nuclear medium by the perturbative expansion method is proposed as a
 natural extension of the Heavy Baryon Chiral Perturbation Theory
 (HBChPT).
Treating the Fermi momentum scale as a separate scale of the system,
we get an improved convergence and the conceptually clear
interpretation. We compute the pion decay constant and the pion
velocity in the nuclear medium, and find their characters different
from what the usual HBChPT predicts. We also obtain the Debye
screening scale at the normal nuclear matter density, and the damping
scale of the pion wave. Those results indicate that the present
theory, albeit its improvement over the HBChPT, has the limitation yet
to go over to the medium of about 1.3 times of normal matter density
due to the absence of the intrinsic density dependence of the
coupling constants. We discuss how we overcome this limitation in
terms of the renormalization method.

\end{abstract}

\renewcommand{\thefootnote}{\#\arabic{footnote}}
\setcounter{footnote}{0}

\section{Introduction}

Low-energy nuclear processes that involve
a few nucleons
can be described in a systematic and consistent way,
thanks to the
heavy-baryon chiral perturbation theory (HBChPT)
which has been initiated by the pioneering work of
Weinberg~\cite{weinberg90}.
This noble approach
has been applied
to many low-energy processes with great successes;
see, for example, Refs. \cite{handbook,force,excs,hep}.

Motivated with these successes,
many already applied the chiral perturbation theory to
dilute systems~\cite{pjm, hammer,oller,MOW,fraga} and also to systems around the saturation density~\cite{lutz,weise}.
When the density and the corresponding Fermi momentum $p_F$
increases,
however,
the convergence of free-space HBChPT becomes questionable.
This is because HBChPT is a derivative expansion scheme with the
expansion parameter $Q/\Lambda_\chi$; $Q$ stands for the
typical momentum scale (and/or the pion mass) of the process, and
$\Lambda_\chi\sim 4\pi f_\pi\simeq 1$ GeV is the chiral scale.
In nuclear matter,
$Q\sim p_F$ even for small fluctuations
whose typical energy-momentum scale is much smaller than $p_F$.
For such fluctuations, the contribution of the nucleons whose
momenta far from the Fermi surface is however irrelevant. It is
thus desirable to build an effective field theory (EFT) by
integrating out such ``{\it massive} degrees of
freedom'', $|\vec{p}| - p_F > \Lambda$, where $\Lambda$ is the
cutoff of the theory.
The resulting EFT has two expansion parameters,  $Q_{ext}/\Lambda_\chi$
and $p_F/\Lambda_\chi$,
where $Q_{ext}$ represents the
energy-momentum scale of the external probes
and/or the pion mass,
$Q_{ext} \le \Lambda$.
The resulting EFT is thus meaningful only when
$Q_{ext} < \Lambda < p_F$ and $p_F < \Lambda_\chi$.
These expansion parameters will emerge naturally in the power counting
rule of the matrix elements of the physical processes, and the
perturbative expansion will follow this counting rule as is the case in
the HBChPT where only the expansion scale is $Q_{ext} < \Lambda_\chi$.
As we shall see later, the new expansion scale $p_F < \Lambda_\chi$ could be threatened to approach to one when the density of the nuclear medium becomes large. But
as we shall show later, the power counting rule shows that the matrix elements depends on the $p_F/\Lambda_\chi$ to twice of the number of
loops that consist only of nucleon lines. This guarantees the rapid convergence in the dilute situation.
In this work, however, we limit ourselves to a modest case where
$Q_{ext} \ll p_F \ll \Lambda_\chi$,
which allows the expansion both in $Q_{ext}/\Lambda_\chi$ and in $p_F/\Lambda_\chi$.
This is the region that can be treated by the HBChPT.
However, the EFT that we are developing here has two important advantages over the
HBChPT.
Firstly, the EFT has only relevant degrees of freedom
(integrating out the irrelevant ones), which is the key aspect of
EFTs that enables us to describe the nature systematically and consistently
in a transparent manner.
Secondly, since the convergence with respect to  powers of $Q_{ext}/\Lambda_\chi$ is
much faster than
that of $p_F/\Lambda_\chi$,
substantial reduction of the relevant diagrams
can be achieved.

The coefficients of the resulting EFT Lagrangian,
{\it i.e.}, low-energy constants (LECs),
will be in general
different from those in the vacuum and must be determined from QCD.
At this moment, we do not have any suitable ways to match our EFT
directly with QCD.
Instead we
fix them by
matching the EFT with the HBChPT;
the Fermi momentum of the matching point should be
upper-bounded by the validity region of HBChPT and lower-bounded
by that of our EFT.
Once the LECs are fixed, we determine the $p_F$-dependences of the
physical quantities by including the (nucleon-)loop corrections.

Before going further, we acknowledge that the idea of building EFT
near the Fermi surface is not new at all. For example, the
familiar Landau-Migdal Fermi theory is one of such. The work of
Schwenk {\it et al}.~\cite{schwenk}
also is worthy to be noted, where the notion
of the $V_{low-k}$~\cite{bogner} has been applied to derive the parameters of
the Landau-Migdal theory. Especially our work shares much in
common with the high density effective theory (HDET)
developed by Hong~\cite{Hong} and by Sch\"afer~\cite{schaefer}, which is an EFT for quark matter
at high-dense region where the matching with the (perturbative)
full QCD is possible. Compared to the HDET, our prime
interests lie in the intermediate density (rather than the
asymptotic) region where
the (almost massless) Goldstone bosons exist and the relevant
Fermion degrees of freedom are still nucleons. However, our
approach may extend its applicability to the region where the
chiral quarks are the relevant degrees of freedom.
Our present study might be relevant for the recent pionic atom
experiment at GSI with heavy nuclei, where the pion decay constant
is interpreted to decrease that may signal the partial chiral
symmetry restoration in medium~\cite{yamazaki}.

We present in this paper our first step to make the in-medium EFT.
Our explicit degrees of freedom are
nucleons near the Fermi surface,
$|\vec{p}|-p_F \leq \Lambda$ and the
Goldstone bosons.
As is claimed recently by one of us (MH)~\cite{HKR}, the role
played by vector mesons at high density should not be
underestimated to realize the important phase of matter at the
transition, the Vector Manifestation (VM)~\cite{HY:VM,HY:PRep}.
Especially, we believe it is very important to understand how this phase
is connected from the normal nuclear matter. Our model indeed
stems from the purpose of its realization. In the present report,
we are however much modest to consider only the pion degrees of
freedom. The underlying rationale lies in the decimation process
at hand. Our theory will allow the shallow excitation nearby the
Fermi surface up to, say, the pionic mass scale. Those modes
beyond it will be integrated out from the beginning. In a sense,
our model will meet the limit of the applicability if the actual
mass of certain integrated-out degrees of freedom becomes lighter
as the density increases. This phenomena will emerge in our EFT
when the parameters do not satisfy the {\it naturalness}. In this
respect, it will be sought in the forthcoming work to include the
chiral partner of pion in the dense limit.

It is worth mentioning some of our ans\"atze.
We are dealing with a dilute system in which loop corrections
to the 4-Fermi type interaction are suppressed by some powers
of $p_F/\Lambda_\chi$.
Furthermore,
the most important channel is assumed to be the
forward scattering one as
in the Fermi liquid~\cite{shankar}, which builds Landau's Fermi
liquid theory~\cite{migdal}.
In the present analysis, we neglect BCS
phenomena. The incorporation of the BCS channel is postponed to
the next step.

This paper is organized as follows:
We give a brief illustration to define the
field expressing the fluctuation mode of the nucleon
in section~\ref{sec:CQF} and
list first few terms in section~\ref{list}.
In section~\ref{pcount}, we build the power counting rule
for our effective Lagrangian and list more higher order
terms.
The in-medium modification of the vector
correlator is studied shortly in section~\ref{vcorr}.
In section~\ref{sec:CPP}, we study the corrections to the
pion decay constants and the pion velocity using our EFT.
The bare parameters for the decay constants will be determined
through the matching between our EFT and the HBChPT.
In section~\ref{mass}, we consider the mass(-like) terms and check
the change in quark condensate due to
mass term in our theory. We
summarize our works in section~\ref{disc}.

\section{Nucleon and its fluctuation fields}
\label{sec:CQF}

In this section, we will define the field corresponding to soft
modes of the fluctuation around the Fermi surface and obtain the
higher order correction terms in the present model scheme.

Let us begin with the free Lagrangian for the nucleon without
interaction at non-zero density:
\begin{eqnarray}
{\cal L} = \bar{\psi}
\left( i \partial_\mu \gamma^\mu - m_N + \mu \gamma_0 \right)
\psi
\ ,
\label{Lag 0}
\end{eqnarray}
where $\mu$ is the baryon chemical potential and $m_N$ the nucleon
mass. Fermi momentum $\vec{p}_F$ is related to the mass $m_N$ and
the chemical potential $\mu$ as in
\begin{equation}
p_F \equiv \left\vert \vec{p}_F \right\vert
= \sqrt{ \mu^2 - m_N^2 } \ .
\end{equation}
In the dense medium of interacting nucleons, we should use the
in-medium nucleon mass,
say $m_N^{\ast}$, instead of the one in vacuum~\cite{Brown-Rho:91}.
But for the notational simplicity,
we refer to $m_N$ as the in-medium
mass $m_N^\ast$ throughout this paper.

In order to consider the fluctuation mode around the Fermi surface, we
introduce the following field $\psi_1$:
\begin{equation}
\psi(x) = \sum_{\vec{v}_F}
  e^{i \vec{p}_F \cdot \vec{x}} \, \psi_1(x;\vec{v}_F) \ ,
\label{def psi'}
\end{equation}
where the Fermi velocity $\vec{v}_F$ is related to the Fermi
momentum
$\vec{p}_F$ as
\begin{equation}
\vec{v}_F \equiv \frac{1}{\mu} \, \vec{p}_F \ .
\label{vF}
\end{equation}
Simply the sum of $\vec{v}_f$ means the angle sum on the Fermi
surface.
By using the field $\psi_1(x;\vec{v}_F)$ in Eq.~(\ref{def psi'}),
the Lagrangian in
Eq.~(\ref{Lag 0}) is rewritten as
\begin{equation}
{\cal L} =
\sum_{\vec{v}_F}
\bar{\psi}_1
\left(
  i \partial_\mu \gamma^\mu - \vec{p}_F \cdot \vec{\gamma}
  - m_N + \mu \gamma_0 \right)
\psi_1
\ ,
\label{Lag 1}
\end{equation}
where $\vec{\gamma} = (\gamma^1, \gamma^2, \gamma^3)$.
To delete off-diagonal components of the $\gamma$ matrices
between $\bar{\psi}_1$ and $\psi_1$ we introduce
the following unitary transformation:
\begin{equation}
\psi_1(x;\vec{v}_F) = U_F(\vec{v}_F) \, \psi_2(x;\vec{v}_F) \ ,
\label{def psi}
\end{equation}
where
\begin{equation}
U_F(\vec{v}_F) \equiv
\cos \theta -
\frac{\vec{v}_F\cdot\vec{\gamma}}{ \bar{v}_F } \, \sin \theta
=
\cos \theta -
\frac{\vec{p}_F\cdot\vec{\gamma}}{ p_F } \, \sin \theta
\ ,
\end{equation}
with $\bar{v}_F = |\vec{v}_F|$ and $\theta$ satisfying
\begin{equation}
\cos \theta = \sqrt{ \frac{\mu+m_N}{2\mu} } \ ,
\quad
\sin \theta = \sqrt{ \frac{\mu-m_N}{2\mu} } \ .
\end{equation}
By using the field $\psi_2(x;\vec{v}_F)$ in Eq.~(\ref{def psi}),
the Lagrangian in Eq.~(\ref{Lag 1}) is rewritten as
\begin{eqnarray}
{\cal L} =\sum_{\vec{v}_F} \bar{\psi}_2 \Biggl[
  i \partial_0 \gamma^0 + i \partial_j v_F^j
  + i \partial_j
    \left( \gamma_\perp^j + \frac{m_N}{\mu} \gp^j \right)
  + \mu \gamma_0 - \mu
\Biggr]
\psi_2
\ ,
\label{Lag 2}
\end{eqnarray}
where
summations over $j$ are implicitly taken and
$v_F^j$ denote components of $\vec{v}_F$ as $\vec{v}_F =
(v_F^1, v_F^2, v_F^3)$. From now on, the indices $i$ and $j$
run over $1$, $2$ and $3$.
Two $\gamma$ matrices representing the
parallel and the orthogonal components to the Fermi velocity,
$\gamma_\parallel^j$ and $\gamma_\perp^j$, are defined by
\begin{eqnarray}
\gp^j &\equiv&
\frac{1}{\bar{v}_F^2} v_F^j (\vec{v}_F\cdot\vec{\gamma})
\ ,
\label{def gam para}
\\
\gamma_\perp^j &\equiv&
\gamma^j - \frac{1}{\bar{v}_F^2} v_F^j (\vec{v}_F\cdot\vec{\gamma})
\ .
\label{def gam perp}
\end{eqnarray}
Now, we decompose the $\psi_2$ field into two parts as
\begin{equation}
\psi_{\pm}(x;\vec{v}_F) \equiv P_{\pm} \,
\psi_2(x;\vec{v}_F)
\ ,
\end{equation}
where the projection operators $P_{\pm}$ are defined as
\begin{equation}
P_{\pm} \equiv \frac{ 1 \pm \gamma_0}{2} \ .
\label{def proj}
\end{equation}
$\psi_+$ and $\psi_-$ represent particle and antiparticle
fields in low energy, respectively.
By using the fields $\psi_{\pm}(x;\vec{v}_F)$, the Lagrangian
in Eq.~(\ref{Lag 2}) is rewritten as
\begin{eqnarray}
{\cal L} &=& \sum_{\vec{v}_F}
\biggl[
\bar{\psi}_{+} \, V_F^\mu \, i \partial_\mu \, \psi_+
+
\bar{\psi}_{-}
\left( - \tilde{V}_F^\mu \, i \partial_\mu - 2 \mu \right)
\psi_-
\nonumber\\
&&
{}+
\bar{\psi}_{-}
\left(
    \gamma_\perp^j + \frac{m_N}{\mu} \, \gp^j
\right)
\, i \partial_j
\psi_+
+
\bar{\psi}_{+}
\left(
    \gamma_\perp^j + \frac{m_N}{\mu} \, \gp^j
\right)
\, i \partial_j
\psi_-
\biggr]
\ ,
\label{Lag 3}
\end{eqnarray}
where
\begin{equation}
V_F^\mu = (1, \vec{v}_F) \ , \quad
\tilde{V}_F^\mu = (1, - \vec{v}_F) \ .
\label{def VF barVF}
\end{equation}
The Lagrangian (\ref{Lag 3}) shows that the field $\psi_+$
expresses the fluctuation mode of the nucleon around the Fermi
surface, and the field $\psi_-$ corresponds to the anti-nucleon
field carrying the effective mass of $2 \mu$ which is large
compared with the momentum scale we are considering. Then, we can
solve out the $\psi_-$ field in studying the low-momentum
region to obtain the Lagrangian including the $\psi_+$ field. It
is to be noted that  $\partial_0 \, \psi_+$ is of same scale as
$\partial_i \, \psi_+ $, thanks to the projection.

In the next section, we will construct the general Lagrangian in
terms of the $\psi_+$ field. Here to have an idea on the higher
order terms, we solve out the $\psi_-$ field in the Lagrangian
(\ref{Lag 3}) at tree level. Using equation of motion for the
$\psi_-$ field
\begin{equation}
\psi_-=
  \frac{\gn^j+\frac{m_N}{\m}\gp^j}{i\tilde{V}_F^\mu \partial_\mu+2\m}
i\del_j\psi_+\ ,
\end{equation}
we obtain
\begin{eqnarray}
{\cal L} &=& \sum_{\vec{v}_F} \Biggl[ \bar{\psi}_{+} \, V_F^\mu \, i
\partial_\mu \, \psi_+ -\bar\p_+\,
\frac{(\tilde{\g}^j\del_j)^2}{2\m} \sum_{n=0}^{\infty}\left(
-\frac{i\tilde{V}_F^\m\del_\m}{2\m} \right)^n\p_+ \Biggr],
\label{free exp}
\end{eqnarray}
with $\tilde{\g}=\gn+\frac{m_N}{\m}\gp$.
This shows that the terms in the right-hand-side (RHS) other than
the first term are suppressed by $Q/\mu$,
where $Q$ denotes the typical momentum scale we are
considering.
Therefore, the first term is the leading order term, which leads to
the following propagator for the fluctuation field
$\psi_+$~\cite{Hong}:
\begin{equation}
i S_F(p) = \frac{1}{ - V_F \cdot p - i \epsilon p_0 }
\, P_+ \ ,
\end{equation}
where $\epsilon \rightarrow + 0$
and $P_+$ is the projection operator defined in Eq.~(\ref{def proj}).
Other terms are regarded as the
higher order terms.
Thus,
the leading correction to the first term of
the RHS is already suppressed considerably at any density.

\section{Effective Lagrangian}
\label{list}

In this section, we will present our effective Lagrangian.
As stressed in Introduction,
we are considering the moderate density region.
So the mesons other than the pion are expected to be
still heavier than the pion, i.e.,
the pion is the only light mesonic degree of freedom
realized as the pseudo Nambu-Goldstone (NG) boson associated
with the spontaneous chiral symmetry breaking.
In the present analysis, therefore,
we build an effective Lagrangian including
the field expressing the fluctuation of the nucleon around the
Fermi surface, $\psi_+$, together with the NG
boson field, based on the non-linear realization of chiral
symmetry by assuming that all the other mesons are integrated out.

As we illustrated in Introduction, we are considering the nucleon
carrying the momentum inside the thin shell of size $\Lambda_{||}$
around the Fermi surface. Then, the momentum of the fluctuation
parallel to the Fermi velocity at the reference point is
restricted to be smaller than $\Lambda_{||}$. Furthermore, we only
consider the situation where the momentum of the fluctuation mode
perpendicular to the Fermi velocity is also smaller than
$\Lambda_{\perp}\sim\Lambda_{||}$ by assuming that effects from
mode carrying large perpendicular momenta are already integrated
out and included in the parameters of the effective Lagrangian.
This idea is based on Landau's Fermi liquid theory where the
forward scattering channel is most important\cite{shankar} and
which works well in nuclear physics~\cite{migdal}. And such a
scheme is being used in HDET for quark matter~\cite{Hong}. Thus,
in the present analysis, we divide the thin shell around the Fermi
surface into small boxes, each of which is characterized
by the point on
the Fermi surface with the characteristic vector, $\vec{v}_F$.
Since the size of each box is $\Lambda_{||}\Lambda_{\perp}^2$,
the momentum of pion is restricted to be small and the
nucleon interacting with pion cannot escape from the box.

\paragraph{Terms without fermion}
\ \par

The pion is introduced as the NG boson associated with the chiral
symmetry breaking of $\mbox{SU}(2)_{\rm L}\times\mbox{SU}(2)_{\rm
R}
 \rightarrow \mbox{SU}(2)_{\rm V}$ through the matrix valued variable
$\xi$
\begin{equation}
\xi = e^{i\pi/F_\pi^t} \ , \label{def xi}
\end{equation}
where $\pi = \sum_{a=1}^{3} \pi_a T_a$ denotes the pion fields,
and $F_\pi^t$ the parameter corresponding to the temporal
component of the pion decay constant.
Here the wave function renormalization of the pion field
is expressed by $F_\pi^t$~\cite{MOW,Sasaki}.
Under the chiral symmetry, this $\xi$ transforms nonlinearly as
\begin{equation}
\xi \rightarrow \xi^{\prime} =
  h(\pi,g_{\rm R},g_{\rm L})\, \xi\, g_R
  = g_{\rm L}\, \xi\, h^{\dag}(\pi,g_{\rm R},g_{\rm L})
\ ,
\end{equation}
where $h(\pi,g_{\rm R},g_{\rm L})$ is an element of
$\mbox{SU}(2)_{\rm V}$ to be uniquely determined. There are two
1-forms constructed from $\xi$:
\begin{eqnarray}
\alpha_{A}^{\mu} &\equiv& \frac{1}{2i} \left(
  {\cal D}^\mu \xi \cdot \xi^\dag
  -
  {\cal D}^\mu \xi^\dag \cdot \xi
\right) \ , \label{al perp}
\\
\alpha_{V}^{\mu} &\equiv& \frac{1}{2i} \left(
  {\cal D}^\mu \xi \cdot \xi^\dag
  +
  {\cal D}^\mu \xi^\dag \cdot \xi
\right) \ , \label{al para}
\end{eqnarray}
where ${\cal D}_\mu \xi$ and ${\cal D}_\mu \xi^\dag$ are defined
as
\begin{eqnarray}
{\mathcal D}_\mu \xi &=& \partial_\mu \xi + i \xi {\mathcal R}_\mu
\ ,
\nonumber\\
{\mathcal D}_\mu \xi^\dag &=&
  \partial_\mu \xi^\dag + i \xi^\dag {\mathcal L}_\mu \ ,
\label{covariant derivative pi}
\end{eqnarray}
with ${\mathcal L}$ and ${\mathcal R}$ being the external gauge
fields for the chiral $\mbox{SU}(2)_{\rm L}\times\mbox{SU}(2)_{\rm
R}$ symmetry. These 1-forms transform as
\begin{eqnarray}
\alpha_{A}^{\mu} &\rightarrow&
  h(\pi,g_{\rm R},g_{\rm L})\, \alpha_{A}^{\mu} \,
  h^\dag(\pi,g_{\rm R},g_{\rm L})
\ ,
\nonumber\\
\alpha_{V}^{\mu} &\rightarrow&
  h(\pi,g_{\rm R},g_{\rm L})\, \alpha_{V}^{\mu} \,
  h^\dag(\pi,g_{\rm R},g_{\rm L})
  - \frac{1}{i}
  h(\pi,g_{\rm R},g_{\rm L})\, \partial^{\mu} \,
  h^\dag(\pi,g_{\rm R},g_{\rm L})
\ .
\end{eqnarray}
By using these 1-forms the general form of the chiral Lagrangian
including pion fields can be constructed. The Lagrangian at the
leading order of the derivative expansion is given by
\begin{equation}
{\mathcal L}_{A0} = \left[
  \left( F_\pi^t \right)^2 u_\mu u_\nu
  + \left( F_\pi^t F_\pi^s \right)
    \left( g_{\mu\nu} - u_\mu u_\nu \right)
\right] \mbox{tr} \left[
  \alpha_{A}^{\mu} \alpha_{A}^{\nu}
\right] \ ,
\label{Lag A0}
\end{equation}
where $u_\mu = (1,\vec{0})$ indicates the rest frame of medium,
and $F_\pi^t$ and $F_\pi^s$ denote the parameters corresponding to
the temporal and spatial pion decay constants.
In our notation $\langle 0 |\partial_\mu A^\mu |\pi \rangle\propto F_\pi^t
p_0^2-F_\pi^s\vec{p}^2 = 0$ for on-shell pions by axial vector
current conservation~\cite{PT:96} so
the pion velocity is expressed as $V_\pi^2 = F_\pi^s/F_\pi^t$.
It should be noted that
since $F_\pi^t$ is the wave function renormalization constant
of the pion field, the chiral symmetry breaking scale is
characterized by
\begin{equation}
 \Lambda_\chi \sim 4\pi F_\pi^t .
\end{equation}

\paragraph{Fermion kinetic term}
\ \par

Now, let us consider the terms of the effective Lagrangian
including the field for expressing the fluctuation mode of the
nucleon around the Fermi surface. In the following, for notational
simplicity, we use $\Psi$ for expressing the fluctuation mode
which was expressed by $\psi_+$ field in the previous section.
Under the chiral symmetry the fluctuation mode $\Psi$ transforms
as
\begin{equation}
\Psi \rightarrow h(\pi,g_{\rm R},g_{\rm L})\, \Psi \ , \label{Psi
trans}
\end{equation}
where $h(\pi,g_{\rm R},g_{\rm L}) \in \mbox{SU}(2)_{\rm V}$. Then
the kinetic term of the $\Psi$ field is expressed as
\begin{equation}
{\cal L}_{\rm kin} = \sum_{\vec{v}_F} \bar{\Psi} \, V_F^\mu \, i
D_\mu \Psi \ , \label{Lag kin}
\end{equation}
where the covariant derivative is defined by
\begin{equation}
D_\mu \Psi =
  \left( \partial_\mu - i \alpha_{V\mu} \right) \Psi
\ , \label{cov der}
\end{equation}
with $\alpha_{V\mu}$ defined in Eq.~(\ref{al para}).

\paragraph{Interactions of fermions with mesons}
\ \par

We consider the interactions of $\Psi$ and $\bar{\Psi}$ to the
$\pi$ field. The possible forms of the interaction among $\Psi$,
$\bar{\Psi}$ and $\alpha_{A }^{\mu}$ defined in
Eq.~(\ref{al perp})
are given by
\begin{eqnarray}
{\cal L}_{A} &=&  \sum_{\vec{v}_F}
 \Bigl[
  i\, \kappa_{A0}
  \bar{\Psi} (\vec{v}_F\cdot \vec{\gamma}) \gamma_5
  {\alpha}_{A}^0 \Psi
  + i \, \kappa_{A\parallel}
  \bar{\Psi} \gamma_{\parallel}^i \gamma_5 {\alpha}_{A i} \Psi
  + i \, \kappa_{A\perp}
  \bar{\Psi} \gamma_{\perp}^i \gamma_5 {\alpha}_{A i} \Psi
 \Bigr]
\ , \label{Lag A}
\end{eqnarray}
where $\kappa_{A0}$, $\kappa_{A\parallel}$ and $\kappa_{A\perp}$
are dimensionless real constants, and $\gamma_{\parallel}^i$ and
$\gamma_{\perp}^i$ are defined in Eqs.~(\ref{def gam perp}) and
(\ref{def gam para}).~\footnote{%
  It may be useful to compare the interaction terms
  in Eq.~(\ref{Lag A}) with those reduced from
  a simple interaction among pion and nucleons given by
  \begin{displaymath}
    i g_A \bar{\psi} \gamma^\mu \gamma_5 \alpha_{A\mu} \psi\ .
  \end{displaymath}
  By performing the same procedure as in section~\ref{sec:CQF},
  the above term is reduced to
  \begin{displaymath}
    i g_A \bar{\Psi} \left[
      (\vec{v}_F\cdot\vec{\gamma}) \gamma_5 \alpha_{A0}
      + \left(
        \frac{m_N}{\mu} \gamma_\perp^i + \gamma_{\parallel}^i
      \right)
      \gamma_5 \alpha_{A i}
    \right] \Psi
  \ ,
  \end{displaymath}
  where $\gamma_\perp^j$ and $\gamma_\parallel^j$ are
  defined in Eqs.~(\ref{def gam perp}) and (\ref{def gam para}).
  Then, when we match the interaction terms in Eq.~(\ref{Lag A})
  with the above terms, we obtain
  the following matching conditions:
  \begin{displaymath}
    \kappa_{A0} = \kappa_{A\parallel} = g_A \ ,
    \quad \kappa_{A\perp} = \frac{m_N}{\mu} g_A \ .
  \end{displaymath}
\label{foot kappa}
}
Note that $\bar{\Psi} \alpha_{A}^0 \Psi$,
$\bar{\Psi} v_F^i \alpha_{A i} \Psi$ and $\bar{\Psi} \left[ v_F^i
\gamma_i \,,\, \alpha_A^j \gamma_j \right] \Psi$ are prohibited by
the parity invariance.

\paragraph{Four Fermi interaction}
\ \par

For constructing four-Fermi interactions, the following relations
are convenient:
\begin{eqnarray}
&& \bar{\Psi} \gamma_0 \Psi = \bar{\Psi} \Psi \ , \quad \bar{\Psi}
\gamma_i \Psi = 0 \ ,
\nonumber\\
&& \bar{\Psi} \gamma_5 \Psi = 0 \ , \quad \bar{\Psi} \gamma_5
\gamma_0 \Psi = 0 \ ,
\nonumber\\
&& \bar{\Psi} \left[ \gamma_0 \,,\, \gamma_i \right] \Psi = 0 \ ,
\quad
\bar{\Psi} \left[ \gamma_i \,,\, \gamma_j \right] \Psi =
2 i \varepsilon_{ijk} \, \bar{\Psi} \gamma_5 \gamma^k \Psi
\ ,
\nonumber\\
&& \bar{\Psi} \left[ \gamma_i \,,\, \gamma_j \right] \gamma_5 \Psi
= 0 \ , \quad \bar{\Psi} \left[ \gamma_0 \,,\, \gamma_i \right]
\gamma_5 \Psi = 2 \bar{\Psi} \gamma_i \gamma_5 \Psi \  \label{rel
bP P}.
\end{eqnarray}

Although the excitation momentum of the nucleon included in each
pair of particle and hole is restricted to be small in the present
analysis, Fermi velocity of one pair can be different from that of
another pair. Thus, using the relation in Eq.~(\ref{rel bP P}), we
can write the four-fermion interactions as
\begin{eqnarray}
{\cal L}_{4F} &=& \frac{F_S}{(F_\pi^t)^2} \left( \sum_{\vec{v}_F}
\bar{\Psi} \Psi \right)^2 +\frac{F_A}{(F_\pi^t)^2}
  \left( \sum_{\vec{v}_F} \bar{\Psi} \gamma_\mu \gamma_5 \Psi \right)^2
\nonumber\\
&&{} +\frac{F_T}{(F_\pi^t)^2}
  \left(
    \sum_{\vec{v}_F} \bar{\Psi} V_F^\mu \gamma_\mu \gamma_5 \Psi
  \right)^2
+\frac{G_S}{(F_\pi^t)^2}
  \left( \sum_{\vec{v}_F} \bar{\Psi}\vec\tau \Psi \right)^2
\nonumber\\
&&{}+\frac{G_A}{(F_\pi^t)^2}
  \left(
    \sum_{\vec{v}_F} \bar{\Psi} \gamma_\mu \gamma_5\vec\tau \Psi
  \right)^2
+\frac{G_T}{(F_\pi^t)^2}
  \left(
    \sum_{\vec{v}_F} \bar{\Psi} V_F^\mu \gamma_\mu \gamma_5\vec\tau
    \Psi
\right)^2 \ . \label{4Fermi 2}
\end{eqnarray}
where $\vec{\tau}/2$ is the generator of SU(2).
It should be noticed that the above four-Fermi interactions
allow two pairs of particle and hole, each of which resides in
different box.  Here we introduce the dimensionless parameters by
rescaling the dimensionful parameters by the temporal component of
the bare pion decay constant. This is natural since we are
constructing the EFT in the chiral broken phase~\cite{manohar}.

\section{Power counting}
\label{pcount}

In this section, we first present the power counting theorem in our
EFT,
which provides that the terms listed in the
previous section are actually the ones of leading order.
Then, we provide the next order terms
according to the power counting.

\subsection{Power counting}

Let us consider the matrix element $M$ with $E_N$ external
nuclear lines, $E_\pi$ external $\pi$ lines and
$E_E$ lines corresponding to the external fields
such as ${\mathcal L}_\mu$ and ${\mathcal R}_\mu$ in
Eq.~(\ref{covariant derivative pi}).
Here, for simplicity we assume that all the mesonic external fields
carry the dimension one and that all the nucleonic external fields
the dimension 3/2.
~\footnote{
  We perform the Fourier transformation for each
  operator associated with the external field.
}

To derive a power counting rule for $M$,
we first observe that the pion and
nucleon propagators that carry the momenta of order of $Q$ are of
order of $Q^{-2}$ and $Q^{-1}$, respectively. Furthermore, we can
count loops that consist at least one pion propagator as of order
of $\int d^4 q \sim \pi^2 Q^4$, where the factor $\pi^2$ is due to
the angle integration.
In fact, these
observations can also be applied to free-space HBChPT.
In our EFT, the integral is realized as
\begin{equation}
\int d^4 p = \sum_{{\vec v_F}}
\int d{\vec l}_\perp^2\, \int d l_\parallel^2
\sim 4 \pi p_F^2 \, \int d^2 l_\parallel\ ,
\end{equation}
since the summation
over ${\vec v}_F$ and the integral over the perpendicular space
corresponds to the area of the Fermi surface.
For the integrations over the momentum parallel to the Fermi momentum
$l_\parallel$ and the energy $l_0$,
we assign $\int d l_0 \,d l_\parallel \sim \pi\,
Q^2$, where the overall factor $\pi$ can be understood either by
the angle integration in 2-D space or by the residue calculation
for the integration over $l_0$.

Recalling the fact that each loop
is accompanied by $(2\pi)^{-4}$, we are thus led to the
following equation
for the counting of a Feynman diagram expressing $M$,
\begin{eqnarray}
M \,
\propto Q^{-2
I_\pi-I_N} \cdot \left(\frac{Q^4}{16 \pi^2}\right)^{L-L_N} \,
\left(\frac{Q^2 p_F^2}{4\pi^2}\right)^{L_N} \, Q^{d}\ ,
\end{eqnarray}
where
$I_{\pi}$ ($I_N$) denotes the number of the pion (nucleon)
propagators, $L$ the number of loops, and $d$ is
the sum of the number of derivatives at $i$-th vertex $d_i$.
$L_N$ is the number of fermionic loops, each of which carries
independent Fermi velocity $\vec{v}_F$,
i.e., number of summations over $\vec{v}_F$.
We note that we are working in the chiral limit, so that
the pion mass does not appear in the above formula.
Using
the techniques developed in HBChPT, we can simplify the above
equation as
\begin{eqnarray}
M \,\sim\,
\left(\frac{Q}{\Lambda_\chi}\right)^\nu\,
\left(\frac{2 p_F}{\Lambda_\chi}\right)^{2L_N} \ ,\label{power}
\end{eqnarray}
with~\footnote{%
 There is ambiguity
 for the overall factor, and thus one may replace
 ``$2p_F/\Lambda_\chi$'' by $p_F/\Lambda_\chi$ in Eq.~(\ref{power}).
}
\begin{eqnarray}
\nu&=& 2 -
\left(\frac{E_N}{2} + E_E\right) + 2 (L-L_N) + \sum_i \nu_i\ ,
\nonumber \\
\nu_i &\equiv& d_i + \frac{n_i}{2} + e_i - 2 \ .
\label{powernu}
\end{eqnarray}
Here
$n_i$ ($e_i$) is the number of nucleons (external
fields) at $i$-th vertex.
For the detailed derivation of the above counting rule, we refer
Ref.~\cite{weinberg90, excV, hep}. Note that we have replaced
$4\pi F_\pi^t$ by $\Lambda_\chi$.

It is worth while mentioning on the comparison of our counting
rule with that of the HDET~\cite{schaefer}. HDET has only one
scale $p_F=\mu$ and the quark loop contributions related to
four-quark interaction is of ${\cal O}(1)$ so all such loop
contributions should be summed. However, the same fermionic loop
contributions are of ${\cal O}((2p_F/\Lambda_\chi)^{2L_N})$ in our
theory.  Though $2p_F/\Lambda_\chi$ is not related to $Q$ directly,
it is a small quantity at moderate densities which satisfy
$p_F\ll \Lambda_\chi$. So $L_N$ is also a good counting parameter
and one-fermionic loop calculation is
enough for moderate density regions,
in which
\begin{equation}
Q \, \ll \, p_F \, \ll \, \Lambda_\chi
\end{equation}
is satisfied.
As density increases and
$p_F$ becomes larger,
we should sum the all fermion loop with the same $\nu$
as we should do in HDET.
Furthermore, we may have to include other degrees of freedom
which become light in the high density region.

\subsection{Higher order terms}
\label{more}

One can easily check that all terms listed in the previous section
are the leading order interactions with $\nu_i=0$.
Here we list higher order terms with $\nu_i=1$ for the next order calculations.

\paragraph{Terms with two derivatives}
\ \par

Possible terms with two derivatives are given by~\footnote{%
  Note that the term $\bar{\Psi} \gamma^i D_i \gamma_5 D^0 \Psi$
  is prohibited by the parity invariance.
}
\begin{equation}
\frac{C_1}{4\pi F_\pi^t} \bar{\Psi} D^\mu D_\mu \Psi
+
\frac{C_2}{4\pi F_\pi^t} \bar{\Psi} \left( V_F^\mu D_\mu \right)^2 \Psi
-
\frac{C_3}{4\pi F_\pi^t} \bar{\Psi} D^0 D^0 \Psi
\ ,
\end{equation}
where the covariant derivative acting on $\Psi$ is defined
in Eq.~(\ref{cov der}).~\footnote{
  It should be noted that the free fermion model provides
  $C_1=C_2=C_3=
  4\pi F^t_\pi/\mu\sim 1$.
}
Note that the term
$\bar{\Psi} \left[ \gamma^i \,,\, \gamma^j \right]
\left[ D_i \,,\, D_j \right] \Psi$ exists.
But this is rewritten as
\begin{eqnarray}
\bar{\Psi} \left[ \gamma^i \,,\, \gamma^j \right]
\left[ D_i \,,\, D_j \right] \Psi
=  - i \, \bar{\Psi} \left[ \gamma^i \,,\, \gamma^j \right]
\, F_{ij}(\alpha_{V}) \Psi
\ ,
\label{2D Vij}
\end{eqnarray}
where
\begin{equation}
F_{\mu\nu}(\alpha_{V})
 \equiv \partial_\mu \alpha_{V\nu}
    - \partial_\nu \alpha_{V\mu}
  - i \, \left[
    \alpha_{V\mu} \,,\, \alpha_{V\nu} \right]
\end{equation}
is the field strength defined from the 1-form $\alpha_{V\mu}$.
We will list this term later.

\paragraph{Terms with one meson field and one derivative}
\ \par

Possible terms including one derivative and one
${\alpha}_A$ field are given by
\begin{eqnarray}
&&
\frac{h_A}{4\pi F_\pi^t}
\bar{\Psi} \left( V_F^\mu D_\mu {\alpha}_{A\nu} \right)
  \gamma^\nu \gamma_5 \Psi
+
\frac{\bar{h}_A}{4\pi F_\pi^t}
\bar{\Psi} \left( \tilde{V}_F^\mu D_\mu {\alpha}_{A\nu} \right)
  \gamma^\nu \gamma_5 \Psi
\ ,
\label{hA term}
\end{eqnarray}
where the derivative $D_\mu$ acts on only ${\alpha}_A$ field as
\begin{equation}
D_\mu \alpha_{A\nu} \equiv
 \partial_\mu \alpha_{A\nu}
  - i \left[ \alpha_{V\mu} \,,\, \alpha_{A\nu} \right]
\ .
\end{equation}
Furthermore, as we showed in Eq.~(\ref{2D Vij}), we have
\begin{eqnarray}
i \, \frac{h_{V0}}{4\pi F_\pi^t}
 \bar{\Psi} \left[ \gamma^i \,,\, \gamma^j \right]
\, F_{ij}(\alpha_{V}) \Psi
=
i \, \frac{h_{V0}}{4\pi F_\pi^t}
 \bar{\Psi} \left[ \gamma^\mu \,,\, \gamma^\nu \right]
\, F_{\mu\nu}(\alpha_{V}) \Psi
\ .
\label{Fmn term}
\end{eqnarray}
Note that there are identities:
\begin{eqnarray}
&&
D_\mu {\alpha}_{A\nu} - D_\nu {\alpha}_{A\mu}
= \widehat{\cal A}_{\mu\nu}
\ ,
\label{rel:perp}
\\
&&
F_{\mu\nu}(\alpha_{V})
=
i \left[
  {\alpha}_{A\mu} \,,\, {\alpha}_{A\nu}
\right]
+ \widehat{\cal V}_{\mu\nu}
\ ,
\label{rel:parallel}
\end{eqnarray}
where $\widehat{\cal A}_{\mu\nu}$
and $\widehat{\cal V}_{\mu\nu}$ are defined as
\begin{eqnarray}
&&
\widehat{\cal V}_{\mu\nu} \equiv \frac{1}{2}
  \left[ \widehat{\cal R}_{\mu\nu} + \widehat{\cal L}_{\mu\nu} \right]
=
\frac{1}{2} \left[
  \xi {\cal R}_{\mu\nu} \xi^\dag
  + \xi^\dag {\cal L}_{\mu\nu} \xi
\right]
\ ,
\nonumber\\
&&
\widehat{\cal A}_{\mu\nu} \equiv \frac{1}{2}
  \left[ \widehat{\cal R}_{\mu\nu} - \widehat{\cal L}_{\mu\nu} \right]
=
\frac{1}{2} \left[
  \xi {\cal R}_{\mu\nu} \xi^\dag
  - \xi^\dag {\cal L}_{\mu\nu} \xi
\right]
\ ,
\label{A V def}
\end{eqnarray}
with ${\cal R}_{\mu\nu}$ and ${\cal L}_{\mu\nu}$ being the field
strengths of the external gauge fields ${\mathcal R}_\mu$ and
${\mathcal L}_\mu$.
So the $h_{V0}$-term in Eq.~(\ref{Fmn term}) can be expressed by
the term
\begin{equation}
i \, \frac{\tilde{h}_{V0}}{4\pi F_\pi^t}
 \bar{\Psi} \left[ \gamma^\mu \,,\, \gamma^\nu \right]
\, \widehat{\mathcal V}_{\mu\nu} \Psi
\end{equation}
and the last term in Eq.~(\ref{2-meson}) below.

\paragraph{Terms with two meson fields}
\ \par

Possible terms with including two of $\alpha_{A}$ field
are given by
\begin{eqnarray}
&&
\frac{h_{AA}^t}{4\pi F_\pi^t}
\bar{\Psi} \alpha_{A}^0 \alpha_A^0 \Psi
+
\frac{h_{AA}^s}{4\pi F_\pi^t}
\bar{\Psi} \alpha_{A}^i \alpha_{A i} \Psi
+
\frac{h_{AA}^{\parallel}}{4\pi F_\pi^t}
\bar{\Psi} \left( \alpha_{A i} v_F^i \right)^2 \Psi
+
i \frac{h_{AA}^a}{4\pi F_\pi^t}
\bar{\Psi} \left[ \alpha_A^\mu \gamma_\mu \,,\,
  \alpha_A^\nu \gamma_\nu \right] \Psi
\ .
\label{2-meson}
\end{eqnarray}

\paragraph{Six Fermi interaction}
\ \par

Simplest terms expressing the six Fermi interaction are easily
obtained by multiplying
each term in Eq.~(\ref{4Fermi 2}) by
$\left( \sum_{\vec{v}_F} \bar{\Psi} \Psi \right)$ or
$\left(
 \sum_{\vec{v}_F} \bar{\Psi} V_F^\mu \gamma_\mu \gamma_5 \Psi
\right)$
like
$\frac{F_{6SS}}{4\pi(F_\pi^t)^5} \left( \sum_{\vec{v}_F}
\bar{\Psi} \Psi \right)^3$.
Other possible terms are expressed as
\begin{eqnarray}
&&
\frac{F_{6I}}{4\pi(F_\pi^t)^5}
\, i \varepsilon_{\mu\nu\alpha\beta}\, V_F^\mu \,
\left( \sum_{\vec{v}_F} \bar{\Psi} \gamma^\nu \gamma_5 \Psi \right)
\left( \sum_{\vec{v}_F} \bar{\Psi} \gamma^\alpha \gamma_5 \Psi \right)
\left( \sum_{\vec{v}_F} \bar{\Psi} \gamma^\beta \gamma_5 \Psi \right)
\nonumber\\
&&{}
+
\frac{G_{6I}}{4\pi(F_\pi^t)^5}
\, i \varepsilon_{\mu\nu\alpha\beta}\, V_F^\mu \,
\left( \sum_{\vec{v}_F} \bar{\Psi} \gamma^\nu \gamma_5
  \vec{\tau} \Psi \right)
\left( \sum_{\vec{v}_F} \bar{\Psi} \gamma^\alpha \gamma_5
  \vec{\tau} \Psi \right)
\left( \sum_{\vec{v}_F} \bar{\Psi} \gamma^\beta \gamma_5 \Psi \right)
\ .
\label{6Fermi2}
\end{eqnarray}

\section{Vector current correlator}
\label{vcorr}

We begin to study our EFT with vector current
correlation function.
In the present case, the vector correlator is simply expressed by
the two-point function of the vector external field ${\mathcal V}_\mu$
defined from the external fields ${\mathcal R}_\mu$ and
${\mathcal L}_\mu$ introduced in Eq.~(\ref{covariant derivative pi})
as
\begin{equation}
{\mathcal V}_\mu = \frac{1}{2}
  \left( {\mathcal R}_\mu + {\mathcal L}_\mu \right)
\ .
\end{equation}
Applying our power counting (\ref{powernu}) to
${\mathcal V}_\mu$-${\mathcal V}_\nu$ two-point function
denoted by $\Pi_{\mathcal V}^{\mu\nu}$
where
$E_N=0$ and $E_E=2$, we have
\begin{eqnarray}
\nu= 2 (L-L_N) + \sum_i \left( d_i + \frac{n_i}{2} + e_i - 2 \right)
\ .
\label{PC 2}
\end{eqnarray}
\begin{figure}[htbp]
\begin{center}
 \includegraphics[width = 12cm]{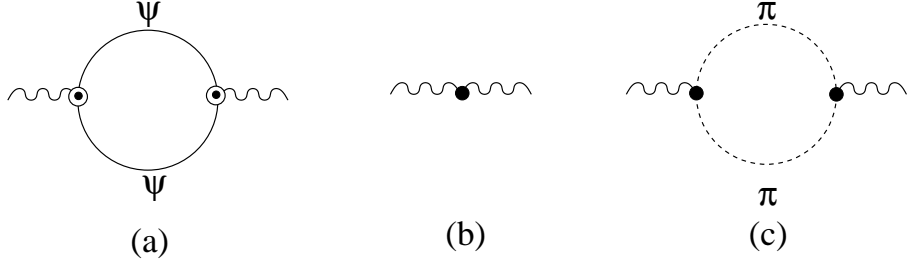}
\end{center}
\caption[]{Diagrams contributing to the
${\mathcal V}_\mu$-${\mathcal V}_\nu$ two-point function:
(a) Fermionic one-loop contribution with $\nu=0$ and $N_L=1$;
(b) Tree contribution with $\nu=2$ and $N_L=0$;
(c) Pion one-loop contribution with $\nu=2$ and $N_L=0$.
Here $\odot$ denotes the vertex with $V_F^\mu$ and $\bullet$
the momentum dependent vertex.
}\label{fig:vv}
\end{figure}
So the leading loop correction is the fermionic
one loop contribution
from the covariant kinetic
term (\ref{Lag kin}), i.e., $\nu =0, L_N=1$
(see Fig.~\ref{fig:vv}(a)):~\footnote{%
 There are no contributions with $\nu = L_N = 0$.
 The contributions with $\nu=2$ and $L_N=0$ are expressed by
 the sum of the tree diagram and the pion one-loop diagram.
 The tree diagram given in Fig.~\ref{fig:vv}(b)
 is constructed from the Lagrangian of the
 following form:
 \begin{eqnarray}
   \left[
     2 z_1^L \, u_\mu u_\alpha g_{\nu\beta}
     + z_1^T \left( g_{\mu\alpha}g_{\nu\beta}
                    - 2 u_\mu u_\alpha g_{\nu\beta} \right)
   \right]
   \,\mbox{tr}
   \left[
      \hat{\mathcal V}^{\mu\nu} \hat{\mathcal V}^{\alpha\beta}
   \right]
 \ ,
 \nonumber
 \end{eqnarray}
 where $\hat{\mathcal V}^{\mu\nu}$ is defined in Eq.~(\ref{A V def}).
 The pion one-loop diagram in Fig.~\ref{fig:vv}(c)
 is constructed from the leading order
 Lagrangian in Eq.~(\ref{Lag A0}).
}
\begin{eqnarray}
\delta_{ab}\Pi_{\mathcal V}^{(1)\mu\nu} (p_0,\vec{p})
&=&
- \mbox{tr} [ T_a T_b ]
\sum_{ \vec{v}_F }
2V_F^\mu V_F^\nu
\int \frac{d^4 l}{i(2\pi)^4}
\frac{1}{
  \left[
    - V_F\cdot ( l - \eta_1 p)
    - i \epsilon (l_0- \eta_1 p_0)
  \right]
}
\nonumber\\
&&
\times
\frac{1}{
  \left[
    - V_F\cdot ( l + \eta_2 p)
    - i \epsilon (l_0 + \eta_2 p_0)
  \right]
}
\ ,
\label{Pi V 0}
\end{eqnarray}
where $u^\mu = (1,\vec{0})$ and $\eta_1$ and $\eta_2$ are constants
satisfying $\eta_1 + \eta_2 = 1$.~\footnote{%
 In the present integration, we cannot make
  the shift of the integration momentum
  in the direction of the Fermi velocity.
  If we were able to make it, we would be able to show that
  the integration in Eq.~(\ref{Pi V 0}) vanishes.
  Then, one might worry that the result of the integral
  depends on the routing of the momentum in each fermion line,
  which would imply that the final result depend on
  $\eta_1$ and $\eta_2$.
  However, as we will show below, the dependence on
  $\eta_1$ and $\eta_2$ will disappear in the final result,
  which indicates that the routing of the loop momentum can be
  done freely at least at one-loop level.
}
In this form the integration over $l_0$ is finite.
Performing the integral over $l_0$, we obtain
\begin{eqnarray}
\Pi_{V}^{(1)\mu\nu} (p_0,\vec{p})
&=&
-
\sum_{ \vec{v}_F }
V_F^\mu V_F^\nu
\frac{1}{ V_F\cdot p + i \epsilon p_0 }
\nonumber\\
&&
\times
\int \frac{d^2 \vec{l}_\perp}{(2\pi)^2}
\int \frac{ d l_{\parallel}}{2\pi}
\left[
  \theta \left( - \vec{v}_F \cdot (\vec{l}-\eta_1 \vec{p}) \right)
  -
  \theta \left( - \vec{v}_F \cdot (\vec{l}+\eta_2 \vec{p}) \right)
\right]
\ ,
\label{pi V 1}
\end{eqnarray}
where $\theta(x)$ represents the step function defined by
\begin{equation}
\theta(x) = \left\{\begin{array}{ll}
  1 \quad & \mbox{for} \ x > 0 \ , \\
  0 & \mbox{for} \ x < 0 \ .
\end{array}\right.
\end{equation}
In Eq.~(\ref{pi V 1})
we replaced the integration over the spatial momentum
as \begin{equation}
\int \frac{d^3\vec{l}}{(2\pi)^3}
= \int \frac{d^2 \vec{l}_\perp}{(2\pi)^2}
\int \frac{ d l_{\parallel}}{2\pi}
\ ,
\label{rep mom int}
\end{equation}
where
\begin{eqnarray}
l_{\parallel} &\equiv& \frac{ \vec{v}_F \cdot \vec{l} }{\bar{v}_F}
\ ,
\nonumber\\
\vec{l}_\perp &\equiv& \vec{l}
 - \vec{v}_F \, \frac{ l_{\parallel}}{\bar{v}_F}
\ ,
\end{eqnarray}
with $\bar{v}_F \equiv \vert \vec{v}_F \vert$.
The integration over $l_{\parallel}$ leads to
\begin{eqnarray}
\Pi_{V}^{(1)\mu\nu} (p_0,\vec{p})
&=&
-
\sum_{ \vec{v}_F }
V_F^\mu V_F^\nu
\int \frac{d^2\vec{l}_{\perp}}{(2\pi)^2}
\frac{1}{2\pi \bar{v}_F}
\frac{\vec{v}_F \cdot \vec{p} }{ V_F\cdot p + i \epsilon p_0 }
\ .
\label{Pi A 1}
\end{eqnarray}
We here consider the integration over $\vec{l}_{\perp}$.
Since
the momentum $\vec{l}_{\perp}$ lies on the Fermi surface,
the combination of
the integration over $\vec{l}_{\perp}$  and the
summation over the Fermi velocity $\vec{v}_F$ implies that
the contribution from overall Fermi surface is included.
Thus, it is natural to make the following replacement
in the above integral:
\begin{equation}
\sum_{ \vec{v}_F }
\int \frac{d^2\vec{l}_{\perp}}{(2\pi)^2}
\Rightarrow
\frac{p_F^2}{\pi} \int \frac{d \Omega_{\vec{v}_F}}{4\pi}
\ .
\label{repl perp}
\end{equation}
Then, we have
\begin{eqnarray}
\Pi_V^{(1)\mu\nu} (p_0,\vec{p})
&=&
- \frac{p_F^2}{2 \pi^2 \bar{v}_F}
\int \frac{d \Omega_{\vec{v}_F}}{4\pi}
V_F^\mu V_F^\nu
\frac{\vec{v}_F \cdot \vec{p} }{ V_F\cdot p + i \epsilon p_0 }
\ .
\label{Pi A 2}
\end{eqnarray}
It should be noticed that
the above contribution
is covariant under the spatial $O(3)$ rotation after the angle
integration is done.
Then, we can decompose this as
\begin{eqnarray}
\Pi_V^{(1)\mu\nu} (p_0,\vec{p})
&\equiv&
u^\mu u^\nu \Pi_{V}^{t}(p_0,\vec{p})
+ \left( g^{\mu\nu} - u^\mu u^\nu \right)
  \Pi_V^{s} (p_0,\vec{p})
\nonumber\\
&&
{}+ P_L^{\mu\nu} \Pi_V^{L} (p_0,\vec{p})
+ P_T^{\mu\nu} \Pi_V^{T} (p_0,\vec{p})
\ ,
\label{Pi V decomp}
\end{eqnarray}
where $P_L^{\mu\nu}$ and $P_T^{\mu\nu}$ are the longitudinal and
transverse polarization
tensors defined by
\begin{eqnarray}
P_L^{\mu \nu}
&=&
- \left( g^{\mu \nu} - \frac{p^\mu p^\nu}{p^2} \right)-
                     P_T^{\mu \nu}
\ ,
\nonumber\\
 P_T^{\mu \nu}
&=&
g^\mu_i
\left( \delta _{ij} - \frac{\vec{p_i}\vec{p_j}}{\bar{p}^2} \right)
g_j^\nu
\ ,
\end{eqnarray}
with $\bar{p} = |\vec{p}|$.
By performing the angle integration in Eq.~(\ref{Pi A 2}),
four components of $\Pi_V^{(1)\mu\nu}$ are expressed as
\begin{eqnarray}
\Pi_V^t(p_0,\bar{p}) &=& 0 \ ,
\nonumber\\
\Pi_V^s(p_0,\bar{p}) &=&
- \frac{p_F^2 \bar{v}_F}{6\pi^2}\ ,
\nonumber\\
\Pi_V^L(p_0,\bar{p}) &=&
- \frac{p_F^2 \bar{v}_F}{2\pi^2} \frac{ p^2 }{ \bar{p}^2 }
  Y(p_0,\bar{p})
\ ,
\nonumber\\
\Pi_V^T(p_0,\bar{p}) &=&
  - \frac{p_F^2 \bar{v}_F }{4\pi^2}
  + \frac{p_F^2}{4\pi^2 \bar{v}_F}
    \frac{ p_0^2 - \bar{v}_F^2 \bar{p}^2 }{ \bar{p}^2 }
    Y(p_0,\bar{p})
\ ,
\end{eqnarray}
where
\begin{eqnarray}
Y(p_0,\bar{p})
&\equiv&
\int \frac{d \Omega_{\vec{v}_F}}{4\pi}
\frac{\vec{v}_F \cdot \vec{p} }{ V_F\cdot p + i \epsilon p_0 }
\nonumber\\
&=&
- 1 + \frac{p_0}{2 \bar{v}_F \bar{p}}
\ln \frac{ p_0 + \bar{v}_F \bar{p} + i \epsilon p_0 }
         { p_0 - \bar{v}_F \bar{p} + i \epsilon p_0 }
\ .
\label{def Y}
\end{eqnarray}

When the external vector current is electromagnetic, the
vector correlator (\ref{Pi A 2}) is the photon self-energy and
represents the screening effects. Using the formula for the static
electric potential energy between two static charges $q_1$ and $q_2$,
\begin{eqnarray}
V(r)=q_1q_2\int\frac{d^3p}{(2\pi)^3}
\frac{e^{i\vec{p}\cdot\vec{r}}}
  {\vec{p}^2+\Pi_V^{00}(p_0\rightarrow 0, \vec{p}^2)}\ ,
\end{eqnarray}
in Coulomb
gauge, the Debye screening mass squared is
\be
 m_D^2=e^2\frac{\mu p_F}{\pi^2}\ .
\label{Debey mass}
\ee
which gives the Debye screening mass about $50$\,MeV at normal nuclear matter density.

The vector correlator $\Pi_V^{(1)}$ in Eq.~(\ref{Pi V decomp})
does not satisfy the current conservation,
$p_\mu \Pi_V^{(1)\mu\nu} \neq 0$.
Like in HDET~\cite{Hong,schaefer},
this is cured by adding a counter term
\begin{eqnarray}
 \La_{CT}
&=& \frac{p_F^2}{4\pi^2 \bar{v}_F}
 \int \frac{d \Omega_{\vec{v}_F}}{4\pi}
\left(\delta_{ij}-\frac{v_F^iv_F^j}{\bar{v_F}^2}\right)
\tr\left[ \alpha_V^\mu \alpha_V^\mu \right]
\nonumber\\
&=&
  \frac{p_F^2}{6\pi^2 \bar{v}_F}
  (g_{\mu\nu} - u_\mu u_\nu) \mbox{tr}
  \left[ \alpha_V^\mu \alpha_V^\nu \right]
 \ .
 \nonumber
\end{eqnarray}
We note that inclusion of the above counter term does not change the
Debey screening mass in Eq.~(\ref{Debey mass}).
As is well known, the Debey mass comes from the longitudinal component
$\Pi_V^L$, which cannot be expressed by a simple local term in the
Lagrangian.
The above counter term gives a contribution to the spatial component
$\Pi_V^s$.

\section{Pion decay constants and pion velocity}
\label{sec:CPP}

In this section we study the
correction to the pion decay constants and pion velocity.
As shown in, e.g., Refs.~\cite{HKR,HY:PRep,Sasaki}, these physical
quantities are extracted from the two-point function of the axial
vector external field ${\mathcal A}_\mu$, of which Feynman diagram is
identical to the one in Fig. 1(a) with the external axial vector fields
instead of ${\mathcal V}_\mu$.

\subsection{Fermionic one-loop correction}
\label{ssec:FOC}

Now, let us consider the fermionic one-loop correction to the
${\cal A}_\mu$-${\cal A}_\nu$ two-point function.
The power counting for the axial vector correlator
is the same as the vector correlator case (\ref{PC 2}).
Then, in the following,
we study the
fermionic one-loop correction to the
${\mathcal A}_\mu$-${\mathcal A}_\nu$ two-point function.
From the interaction term in Eq.~(\ref{Lag A}),
the correction with $\nu = 0$ and $N_L = 1$ is expressed as
\begin{eqnarray}
\delta_{ab}
\Pi_A^{(1)\mu\nu} (p_0,\vec{p})
&=&
-
\mbox{tr} [ T_a T_b ]
\sum_{ \vec{v}_F }
2
\Biggl[
  \kappa_{A0}^2 \bar{v}_F^2 u^\mu u^\nu
  + \kappa_{A\parallel}^2 g^\mu_i g^\nu_j
    \frac{v_F^i v_F^j}{\bar{v}_F^2}
\nonumber\\
&& \quad
  {}+ \kappa_{A\perp}^2  g^\mu_i g^\nu_j
      \left( \delta^{ij} - \frac{v_F^i v_F^j}{\bar{v}_F^2} \right)
  + \kappa_{A0}\kappa_{A\parallel}
      \left( u^\mu g^\nu_j v_F^j + u^\nu g^\mu_j v_F^j \right)
\Biggr]
\nonumber\\
&&
\times
\int \frac{d^4 l}{i(2\pi)^4}
\frac{1}{
  \left[
    - V_F\cdot ( l - \eta_1 p)
    - i \epsilon (l_0- \eta_1 p_0)
  \right]
}
\nonumber\\
&&
\times
\frac{1}{
  \left[
    - V_F\cdot ( l + \eta_2 p)
    - i \epsilon (l_0 + \eta_2 p_0)
  \right]
}
\nonumber\\
&=&
- \delta_{ab}\frac{p_F^2}{2\pi^2 \bar{v}_F}
\int \frac{d \Omega_{\vec{v}_F}}{4\pi}
\Biggl[
  \kappa_{A0}^2 \bar{v}_F^2 u^\mu u^\nu
  + \kappa_{A\parallel}^2 g^\mu_i g^\nu_j
    \frac{v_F^i v_F^j}{\bar{v}_F^2}
\nonumber\\
&& \quad
  {}+ \kappa_{A\perp}^2
      g^\mu_i g^\nu_j\left( \delta^{ij} - \frac{v_F^i v_F^j}{\bar{v}_F^2} \right)
  + \kappa_{A0}\kappa_{A\parallel}
      \left( u^\mu g^\nu_j v_F^j + u^\nu g^\mu_j v_F^j \right)
\Biggr]
\nonumber\\
&&
\times
\frac{\vec{v}_F \cdot \vec{p} }{ V_F\cdot p + i \epsilon p_0 }
\ .
\label{Pi A 2 b}
\end{eqnarray}

The contribution to ${\cal A}_\mu$-${\cal A}_\nu$ two-point function
at leading order with $\nu = N_L = 0$ is expressed as
\begin{eqnarray}
\Pi_A^{(0)\mu\nu} (p_0,\vec{p})
=
\left( F_\pi^t \right)^2 u^\mu u^\nu
+ F_\pi^t F_\pi^s \left( g^{\mu\nu} - u^\mu u^\nu \right)
\ .
\label{Pi A tree}
\end{eqnarray}
As is done for the vector current correlator in
Eq.~(\ref{Pi V decomp}),
it is convenient to
decompose the sum of the above contributions as
\begin{eqnarray}
\Pi_A^{\mu\nu} (p_0,\vec{p})
&\equiv&
\Pi_A^{(0)\mu\nu} (p_0,\vec{p})
+
\Pi_A^{(1)\mu\nu} (p_0,\vec{p})
\nonumber\\
&=&
u^\mu u^\nu \Pi_{A}^{t}(p_0,\vec{p})
+ \left( g^{\mu\nu} - u^\mu u^\nu \right)
  \Pi_A^{s} (p_0,\vec{p})
{}+ P_L^{\mu\nu} \Pi_A^{L} (p_0,\vec{p})
+ P_T^{\mu\nu} \Pi_A^{T} (p_0,\vec{p})
\ .
\nonumber\\
\label{Pi A decomp}
\end{eqnarray}
{}From Eqs.~(\ref{Pi A 2 b}) and (\ref{Pi A tree}),
two components $\Pi_A^t$ and $\Pi_A^s$ are obtained as
\begin{eqnarray}
\Pi_{A}^{t}(p_0,\vec{p})
&=&
(F_\pi^t)^2
+ \frac{p_F^2}{ 2\pi^2 \bar{v}_F}
\left(
  \kappa_{A0}^2 \bar{v}_F^2 - \kappa_{A0} \kappa_{A{\parallel}}
\right)
Y(p_0,\bar{p})
\ ,
\label{Pi At}
\\
\Pi_{A}^{s}(p_0,\vec{p})
&=&
F_\pi^t F_\pi^s
+ \frac{p_F^2}{ 2\pi^2 \bar{v}_F}
\Biggl[
  \frac{1}{3}
    \left( \kappa_{A{\parallel}}^2 - \kappa_{A\perp}^2 \right)
\nonumber\\
&& \qquad\quad
  {}+
  \left\{
    \left(
      \kappa_{A0}\kappa_{A\parallel}
      - \frac{\kappa_{A\parallel}^2 - \kappa_{A\perp}^2}{\bar{v}_F^2}
    \right)
      \frac{p_0^2}{\bar{p}^2}
    - \kappa_{A\perp}^2
  \right\}
  Y(p_0,\bar{p})
\ ,
\label{Pi As}
\end{eqnarray}
where the function $Y(p_0,\bar{p})$ is defined in
Eq.~(\ref{def Y}).

Using the decomposition of $\Pi_A^{\mu\nu}$ in
Eq.~(\ref{Pi A decomp}),
we can construct
the axial vector current correlator as
\begin{equation}
G_A^{\mu\nu}(p_0,\vec{p}) =
P_L^{\mu\nu} G_A^L(p_0,\vec{p}) +
P_T^{\mu\nu} G_A^T(p_0,\vec{p})
\ ,
\end{equation}
where
\begin{eqnarray}
G_A^L(p_0,\vec{p})
&=&
\frac{ p^2 \Pi_A^t \Pi_A^s}
{-\left[ p_0^2 \Pi_A^t - \bar{p}^2 \Pi_A^s \right] }
+ \Pi_A^L
\ ,
\nonumber\\
G_A^T(p_0,\vec{p})
&=&
- \Pi_A^s + \Pi_A^T
\ .
\end{eqnarray}
We define the {\it on-shell} of the pion from the pole of the
longitudinal component $G_A^L$ of axial vector current correlator.
Since $\Pi_A^t$ and $\Pi_A^s$
have imaginary parts, we determine the
pion energy $E$ from the on-shell of the real part by solving the
following dispersion formula:
\begin{eqnarray}
  0
&=&
  \left[
    p_0^2 \, \mbox{Re} \Pi^{t}_A (p_0,\bar{p})
    - \bar{p}^2 \, \mbox{Re} \Pi^{s}_A (p_0,\bar{p})
  \right]_{p_0=E}
\ .
\label{pi on shell}
\end{eqnarray}
The pion velocity is defined by the relation
\be
v_\pi(\bar{p}) \equiv E/\bar{p}\ee
and is then obtained by solving
\begin{equation}
v_\pi^2(\bar{p}) =
\frac{ \mbox{Re} \Pi^{s}_A (p_0=E,\bar{p})}
{ \mbox{Re} \Pi^{t}_A (p_0=E,\bar{p})}
\ .
\label{vpi}
\end{equation}
{}From the expressions of
$\Pi^{t}_A (p_0,\bar{p})$
and $\Pi^{s}_A (p_0,\bar{p})$
in Eqs.~(\ref{Pi At}) and (\ref{Pi As}),
$\mbox{Re} \Pi^{t}_A (p_0=E,\bar{p})
  = \mbox{Re} [f_\pi^t]^2$
and
$\mbox{Re} \Pi^{s}_A (p_0=E,\bar{p})
  = \mbox{Re} [f_\pi^t f_\pi^s]$
are expressed as
\begin{eqnarray}
&&
\mbox{Re} [f_\pi^t]^2
=
\mbox{Re} \Pi^{t}_A (p_0=E,\bar{p})
=
(F_\pi^t)^2
+ \frac{p_F^2}{2 V_{\pi} }
\left(
  \kappa_{A0}^2 \bar{v}_F^2 - \kappa_{A0} \kappa_{A{\parallel}}
\right)
X\left( \frac{\bar{v}_F}{V_{\pi}} \right)
\ ,
\label{Pi At os}
\\
&&
\mbox{Re} [f_\pi^t f_\pi^s]
=
\mbox{Re} \Pi^{s}_A (p_0=E,\bar{p})
=
F_\pi^t F_\pi^s
+ \frac{p_F^2}{ 6\pi^2 \bar{v}_F}
    \left( \kappa_{A{\parallel}}^2 - \kappa_{A\perp}^2 \right)
\nonumber\\
&& \qquad \qquad
{}+ \frac{p_F^2}{2 V_{\pi} }
  \left\{
    V_{\pi}^2
    \left(
      \kappa_{A0}\kappa_{A\parallel}
      - \frac{\kappa_{A\parallel}^2 - \kappa_{A\perp}^2}{\bar{v}_F^2}
    \right)
    - \kappa_{A\perp}^2
  \right\}
  X\left( \frac{\bar{v}_F}{V_{\pi}} \right)
\ ,
\label{Pi As os}
\end{eqnarray}
where $V_{\pi} \equiv \sqrt{ F_\pi^s/F_\pi^t }$
is the bare pion velocity and the function
$X(r)$ is defined by
\begin{equation}
X(r) \equiv
\frac{1}{ \pi^2 r }
\left[
  - 1 + \frac{1}{2 r}
  \ln \left\vert \frac{ 1 + r }{ 1 - r} \right\vert
\right]
\ .
\end{equation}
Note that we have replaced $v_\pi(\bar{p})$ with
$V_{\pi}$ in the one-loop contribution in
Eqs.~(\ref{Pi At os}) and (\ref{Pi As os})
since the difference is of higher order.

The damping factor of the pion, $\gamma(\bar{p})$, can be defined from
the imaginary part of the denominator of the longitudinal axial vector
current correlator as
\begin{equation}
\gamma(\bar{p}) =
\frac{ \bar{p} }{ 2 \mbox{Re} \Pi_A^t(p_0=E,\bar{p}) }
\,
\mbox{Im} \left[
  \Pi_A^t(p_0=E,\bar{p}) - \Pi_A^s(p_0=E,\bar{p})
\right]
\end{equation}
By substituting the expressions of
$\Pi^{t}_A (p_0,\bar{p})$
and $\Pi^{s}_A (p_0,\bar{p})$
in Eqs.~(\ref{Pi At}) and (\ref{Pi As}), this is rewritten as
\begin{eqnarray}
\frac{ \gamma(\bar{p}) }{ \bar{p} }
&=&
\frac{1}{ 2 \mbox{Re} \Pi_A^t(p_0=E,\bar{p}) }
\frac{V_\pi p_F^2 }{ 2 \pi \bar{v}_F^2 }
\,\theta ( \bar{v}_F - V_\pi )
\nonumber\\
&& \qquad
\times
\Biggl[
  \left\{
    \kappa_{A0}^2 \bar{v}_F^2 - \kappa_{A0} \kappa_{A{\parallel}}
  \right\}
  -
  \left\{
    V_\pi^2
    \left(
      \kappa_{A0}\kappa_{A\parallel}
      - \frac{\kappa_{A\parallel}^2 - \kappa_{A\perp}^2}{\bar{v}_F^2}
    \right)
    - \kappa_{A\perp}^2
  \right\}
\Biggr]
\ .
\label{damp}
\end{eqnarray}

\subsection{Determination of parameters}

We should stress that
the parameters of the EFT must be determined from
QCD.
One way to determine them is the Wilsonian
matching~\cite{HY:WM,HY:PRep}, in which the current correlators
in the EFT are matched with those obtained in the operator
product expansion (OPE) around 1\,GeV.
As is stressed in Refs.~\cite{HS:VMT,HKR},
when we make
the matching in the presence of hot and/or dense matter,
the parameters of the EFT generally depend on the
temperature and/or density (the intrinsic temperature
and/or density dependence).
In the present case, however, the cutoff of our EFT is around a
few hundred MeV at most, so that we cannot make the matching
between our EFT and QCD directly.
So, here we consider the matching between our EFT and the
HBChPT to determine
the parameters included in our EFT.
We require that the temporal and spatial pion decay constants
calculated in our EFT agree with those calculated in the HBChPT
at a matching point, say $\rho_M$.

In the HBChPT at one-loop, the temporal and spatial pion decay
constants are expressed as~\cite{MOW}
\begin{eqnarray}
f_\pi^{{\rm(HB)}t} &=& f_\pi
\left[
  1 + \frac{2\rho}{f_\pi^2}
    \left( c_2 + c_3 - \frac{g_A^2}{8m_N} \right)
\right]
\ ,
\label{fpt HB}
\\
f_\pi^{{\rm(HB)}s} &=& f_\pi
\left[
  1 - \frac{2\rho}{f_\pi^2}
    \left( c_2 - c_3 + \frac{g_A^2}{8m_N} \right)
\right]
\label{fps HB}
\end{eqnarray}
where $f_\pi$ is the pion decay constant
at zero density.

The matching conditions are given by
equating Eq.~(\ref{Pi At os}) with Eq.~(\ref{fpt HB})
and Eq.~(\ref{Pi As os}) with Eq.~(\ref{fps HB})
at $\rho = \rho_M$:
\begin{eqnarray}
\left. \mbox{Re}[f_\pi^t]^2 \right\vert_{\rho = \rho_M}
=
\left. \left[ f_\pi^{{\rm(HB)}t} \right]^2
\right\vert_{\rho = \rho_M}
\ ,
\nonumber\\
\left. \mbox{Re}[f_\pi^t f_\pi^s] \right\vert_{\rho = \rho_M}
=
\left. \left[ f_\pi^{{\rm(HB)}t} f_\pi^{{\rm(HB)}s} \right]
\right\vert_{\rho = \rho_M}
\ .
\label{match cond 0}
\end{eqnarray}

The above conditions are not
enough to determine all the parameters, so we adopt the following
ansatz to determine the values of
$\kappa_{A0}$, $\kappa_{A\parallel}$ and $\kappa_{A\perp}$ [see
footnote~\ref{foot kappa}]:
\begin{equation}
\kappa_{A0} = \kappa_{A\parallel} = g_A \ ,
  \quad \kappa_{A\perp} = \frac{m_N}{\mu} g_A \ ,
\label{approx kappa}
\end{equation}
where $g_A$ is the axial coupling.
Here we use the following values of $m_N$ and $g_A$
in the matter free vacuum:
\footnote{
 In the high density region, $m_N$ should
 be replaced with $m_N^{\ast}$.
 Here we consider the density region up until around the normal
 nuclear matter density, so that the vacuum mass will give a good
 approximation.
 The experimental date shows that there is a $10\,\%$ decrease
 in $f_\pi$ due to the in-medium modification~\cite{yamazaki}.
 Thus if $m_N^\ast$ decreases by $10\,\%$,
 our results may be changed by a few \%.
}
\begin{equation}
m_N = 939\, \mbox{MeV}\,, \qquad
g_A = 1.267 \ .
\label{mN,gA}
\end{equation}
In the approximation (\ref{approx kappa}), the temporal and
spatial pion decay constants in Eqs.~(\ref{Pi At os})
and (\ref{Pi As os}) become
\begin{eqnarray}
&&
\mbox{Re} [f_\pi^t]^2
=
(F_\pi^t)^2
- \frac{p_F^2}{2 v_{\pi0} } \frac{m_N^2}{\mu^2} \, g_A^2\,
  X\left( \frac{\bar{v}_F}{v_{\pi0}} \right)
\ ,
\label{Pi At os 2}
\\
&&
\mbox{Re} [f_\pi^t f_\pi^s]
=
F_\pi^t F_\pi^s
+ \frac{p_F^2}{ 6\pi^2 } \bar{v}_F g_A^2
- \frac{p_F^2}{2 v_{\pi0} } \frac{m_N^2}{\mu^2} \, g_A^2\,
  X\left( \frac{\bar{v}_F}{v_{\pi0}} \right)
\ .
\label{Pi As os 2}
\end{eqnarray}
It should be noted that
the damping factor of pion vanishes when we take the ansatz in
Eq.~(\ref{approx kappa}) with $\mu=\sqrt{p_F^2+m_N^2}$:
\begin{equation}
\gamma(\bar{p}) = 0 \ .
\end{equation}

Now, by using the matching conditions in Eq.~(\ref{match cond 0})
together with the ansatz in Eq.~(\ref{approx kappa}), we determine
the values of the parameters $F_\pi^t$ and $F_\pi^s$ for given
matching density $\rho_M$.
We take the values for the low-energy constants in the HBChPT as
$c_2 = 3.2 \pm 0.25\,\mbox{GeV}^{-1}$~\cite{Fettes:1998ud,Fettes:2000xg}
and $c_3 = -4.70 \pm 1.16\,\mbox{GeV}^{-1}$
~\cite{Buttiker:1999ap,Rentmeester:1999vw}.
Substituting these values into Eqs.~(\ref{fpt HB}) and (\ref{fps HB}),
we can reduce them to~\cite{MOW}
\begin{eqnarray}
 &&f_\pi^{{\rm (HB)}t}
 = f_\pi \Biggl[ 1 - \frac{\rho}{\rho_0}
     (0.26 \pm 0.04) \Biggr], \\
 &&f_\pi^{{\rm (HB)}s}
 = f_\pi \Biggl[ 1 - \frac{\rho}{\rho_0}
     (1.23 \pm 0.07) \Biggr],
\end{eqnarray}
where we take the pion decay constant in the vacuum as
$f_\pi = 92.4\,\mbox{MeV}$ and the nucleon mass as Eq.~(\ref{mN,gA}),
and $\rho_0$ denotes the normal nuclear density.
When we take the matching density as $\rho_M / \rho_0 = 0.2, 0.3$ and $0.4$,
the values of $F_\pi^t$ and $F_\pi^s$ are obtained by solving
the matching conditions Eq.~(\ref{match cond 0}) as follows:
\begin{equation}
 (F_\pi^t, F_\pi^s)
  = \left\{
    \begin{array}{@{\,}ll}
    (88.3\,\mbox{MeV},\, 69.6\,\mbox{MeV})
     \quad \mbox{for} \quad \rho_M / \rho_0 = 0.2 \,,
\\
    (87.1\,\mbox{MeV},\, 59.1\,\mbox{MeV})
     \quad \mbox{for} \quad \rho_M / \rho_0 = 0.3 \,,
\\
    (87.3\,\mbox{MeV},\, 50.6\,\mbox{MeV})
     \quad \mbox{for} \quad \rho_M / \rho_0 = 0.4 \,.
    \end{array}
    \right.
\end{equation}

\subsection{Density Dependence of the Pion Decay Constants
 and Pion Velocity}

\begin{figure}
 \begin{center}
  \includegraphics[width = 4.5cm]{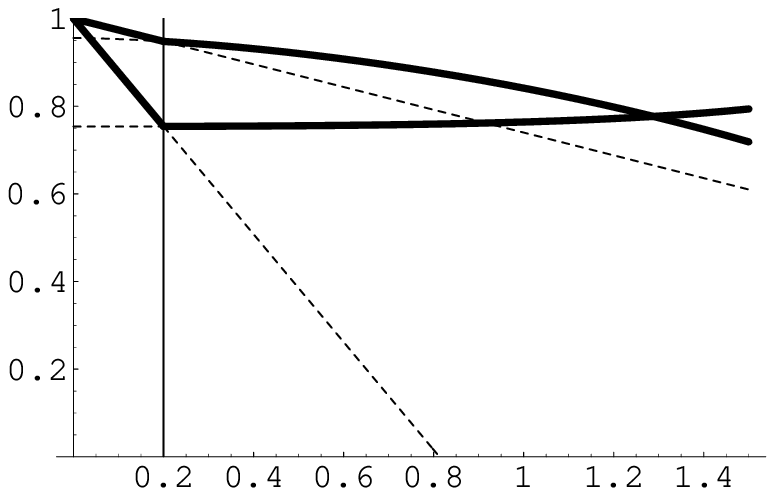}
  \includegraphics[width = 4.5cm]{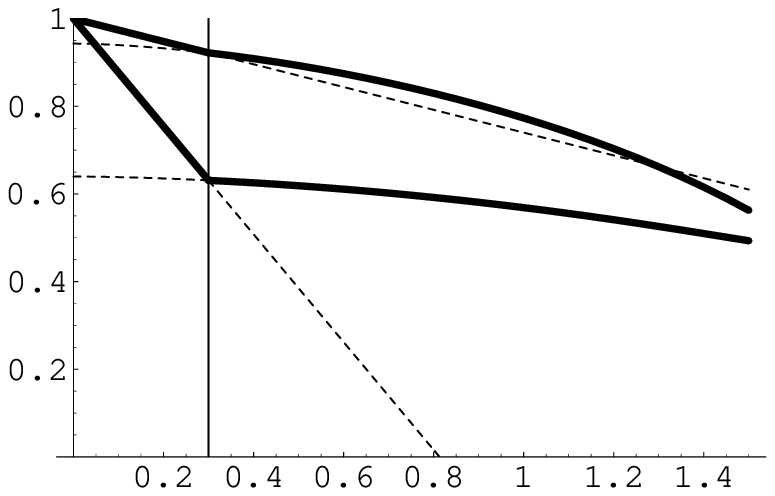}
  \includegraphics[width = 4.5cm]{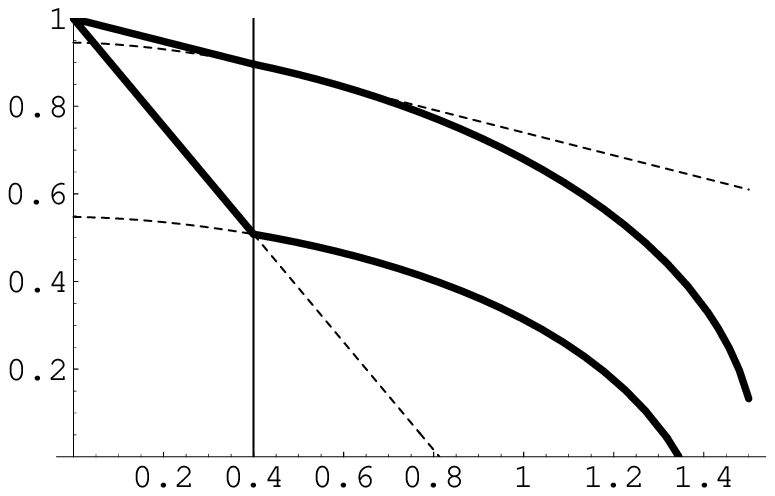}

  (a) \hspace*{3.8cm} (b) \hspace*{3.8cm} (c)
 \end{center}
 \caption{
 Density dependences of the pion decay constants with the matching
 point
 $\rho_M / \rho_0 = 0.2$ (a), $0.3$ (b) and $0.4$ (c).
 The horizontal axis shows the value of $\rho / \rho_0$,
 and the vertical axis the value of the pion decay constants
 normalized by the vacuum value.
 The vertical solid line shows the position of the matching density,
 $\rho_M / \rho_0$.
 In the low-density region, $\rho < \rho_M$,
 the thick solid lines denote $f_\pi^t / f_\pi$ (upper line) and
 $f_\pi^s / f_\pi$ (lower one) obtained from the HBChPT,
 and the dashed lines denote $f_\pi^t / f_\pi$ (upper line) and
 $f_\pi^s / f_\pi$ (lower one) obtained from our EFT.
 In the higher density region, $\rho > \rho_M$,
 we used thick solid lines for the predictions obtained from our EFT
 and dashed lines for those from the HBChPT.
}
 \label{fig:fpi}
\end{figure}

In this subsection,
we study
the pion decay constants and pion velocity in dense matter.
Strictly speaking,
the parameters of the EFT generally have the intrinsic density
dependence.
In our construction of the present EFT,
we expand the Lagrangian by $1/\mu$ and refer to the coefficient
of each power of $1/\mu$ as the parameter.
Then, the density dependences of the parameters are moderate.
As a result, the density dependences of the physical quantities
are dominated by the dense effect from the fluctuating
nucleon
loop in the density region not so much higher than
the matching point $\rho_M \sim (0.2-0.4)\rho_0$.
Based on this,
we use the values of the parameters obtained in the previous
subsection
to obtain the density dependences of the
pion decay constants and pion velocity through the
dense loop correction from the fluctuation mode.
We show the density dependences of the temporal and spatial pion decay
constants in Fig.~\ref{fig:fpi}.
Until the matching density $\rho_M$,
the behavior of physical quantities is described by the HBChPT.
The HBChPT breaks down at the density where $f_\pi^s$ vanishes,
around $0.8 \rho_0$.
Before the HBChPT breaks down, we should switch the theory from the
HBChPT to our EFT in which the dominant density dependences of the
physical quantities come from the nucleon fluctuating near the Fermi
surface.

As one can see from the expressions of Eqs.~(\ref{fpt HB})
and (\ref{fps HB}), the density dependence of the pion decay constants
in the HBChPT is proportional to $p_F^3$.
This is the reflection that the nucleon inside Fermi sphere
contributes to physical quantities.
While the pion decay constants in our EFT have the density dependence
of
$p_F^2$ [see Eqs.~(\ref{Pi At os 2}) and (\ref{Pi As os
2})],~\footnote{%
 The baryon chemical potential $\mu$ and function
 $X(\bar{v}_F / V_\pi)$ have the implicit $p_F$ dependences.
 However these density dependences are small as compared with
 $p_F^2$.
}
which is a consequence that our EFT includes the effects of the
nucleon near Fermi surface.
Thus the density dependence of physical quantities becomes gentle
and then the available region extends as shown in Fig.~\ref{fig:fpi}.

As well as the HBChPT, our EFT also breaks down at the density
where $f_\pi^s$ becomes zero.
Figure~\ref{fig:fpi}(c) shows that the EFT is not applicable at
$\rho \simeq 1.3 \rho_0$.
Even if $f_\pi^s$ is still finite, the EFT is not valid at the density
where $f_\pi^s$ becomes larger than $f_\pi^t$ because of the
inconsistency with causality.
{}From Fig.~\ref{fig:fpi}(a) and (b),
we find that the EFT breaks down at $\rho \simeq 1.3 \rho_0$
and $\rho \simeq 1.8 \rho_0$, respectively.

Here in order to check the availability of our EFT,
we compare one-loop correction with tree contribution.
Looking for the density where the ratio of one-loop correction to
tree contribution becomes $0.5$,
we find such density as follows~\footnote{
 The ratio for $f_\pi^t$ is almost same as that for $f_\pi^s$.
}:
$\rho \simeq 1.2 \rho_0$ (for $\rho_M / \rho_0 = 0.3$) and
$\rho \simeq 0.8 \rho_0$ (for $\rho_M / \rho_0 = 0.4$).
As we mentioned above, Fig.~\ref{fig:fpi}(a) shows that
our EFT breaks down at $\rho \simeq 1.3 \rho_0$.
Then in the case (a) our EFT becomes invalid before perturbative
expansion becomes unreliable [The ratio at $\rho \simeq 1.3 \rho_0$
in this case is $0.2 - 0.3$, which is still small].
Thus in any case our EFT describes physical quantities
until around the normal nuclear density $\rho_0$.

Note that we just calculate $\nu=0$ and $L_N=1$. We need to sum
all the fermion loops with the same $\nu$ as the density
increases. 1.3$\rho_0$ indicates how far we can keep $L_N$
as a counting parameter, not how far we can go with counting
parameter $\nu$. By summing fermion loops with 4-Fermi interactions,
we can formally extend our theory much further.
In the high density region, however, the intrinsic density dependence
of the parameters of the EFT will be important.
It is crucial for studying the density dependences of the physical
quantities that we include the intrinsic density dependences of the
parameters in some way.
Furthermore, we may have to include other degrees of freedom
which becomes light in the high density region.

\begin{figure}
 \begin{center}
  \includegraphics[width = 4.5cm]{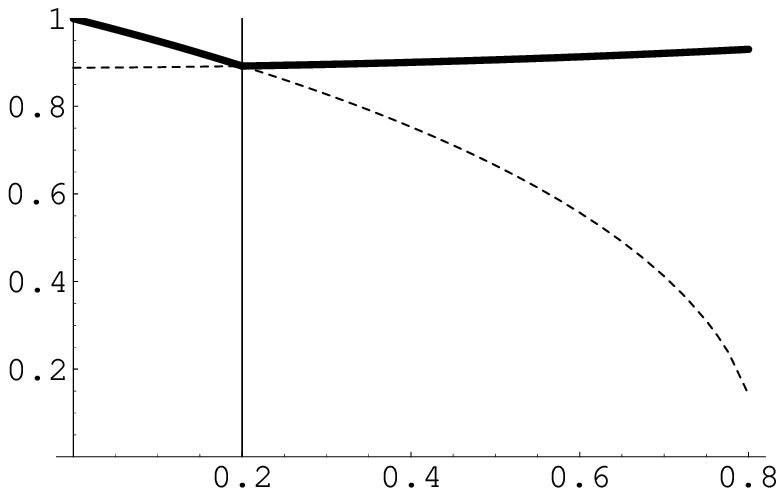}
  \includegraphics[width = 4.5cm]{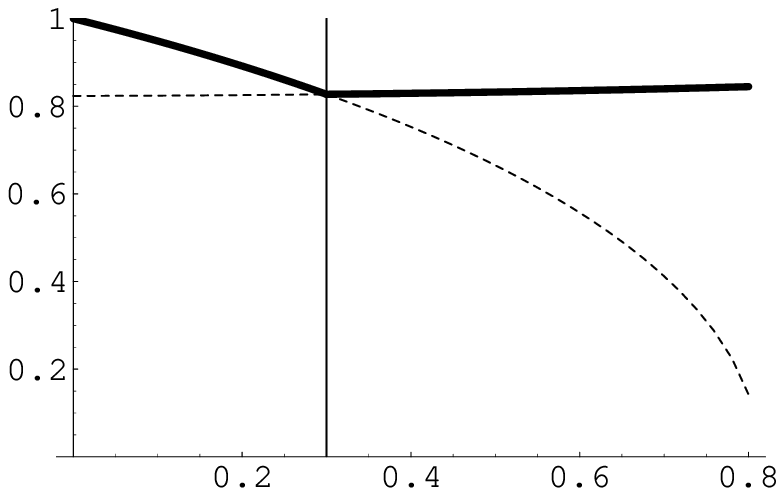}
  \includegraphics[width = 4.5cm]{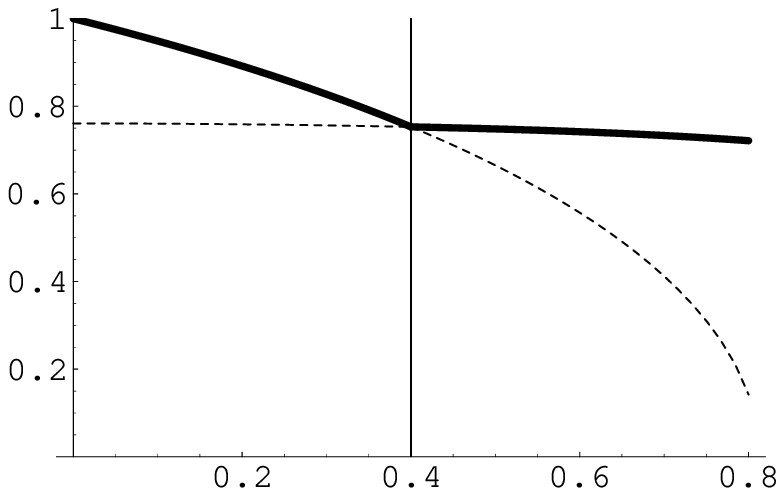}

  (a) \hspace*{3.8cm} (b) \hspace*{3.8cm} (c)
 \end{center}
 \caption{
 Density dependence of the pion velocity with the matching point
 $\rho_M / \rho_0 = 0.2$ (a), $0.3$ (b) and $0.4$ (c).
 The horizontal axis shows the value of $\rho / \rho_0$,
 and the vertical axis the value of the pion velocity.
 The position of the matching density, $\rho_M / \rho_0$,
 is indicated by the vertical solid lines.
 For $\rho < \rho_M$,
 the thick solid line denotes the prediction of the HBChPT
 and the dashed line the one of our EFT.
 For $\rho > \rho_M$, on the other hand,
 we used thick solid line for our EFT
 and dashed line for the HBChPT.
}
 \label{fig:vpi}
\end{figure}

{}From Eq.~(\ref{vpi}) and the above results on $f_\pi^t$ and
$f_\pi^s$,
we can determine the pion velocity $v_\pi$.
Figure~\ref{fig:vpi} shows the density dependence of $v_\pi$.
The pion velocity becomes smaller with increasing density
in low density region.
Since the low-energy constant $c_2$ in the HBChPT is comparable with
tree contribution, the pion velocity largely decreases.
In the region described by our EFT, the pion velocity scarcely change
with respect to density.
This behavior is quite similar to the one predicted by the skyrmion
approach in the moderate density region~\cite{LPMV}.

\section{Mass(-like) terms and condensate}
\label{mass}

\subsection{Two mass terms}

In section 2, we have deleted the original mass term.
But the transformation property of $\Psi$ given in
Eq.~(\ref{Psi trans}) does not prohibit the existence of the
term $\bar{\Psi} \Psi$.
Then, one might think that
we should include the mass term of the fluctuation field $\Psi$
as
\begin{equation}
{\mathcal L}_{\bar{\mu}} = \sum_{\vec{v}_F}
\bar{\mu} \bar{\Psi} \Psi
\end{equation}
where $\bar{\mu}$ denotes the mass parameter.
Since $\bar{\Psi}\Psi = \Psi^\dag \Psi$,
the effect of $\bar{\mu}$ should be included into the definition
of the chemical potential $\mu$ and
this term should be omitted in the Lagrangian.

Furthermore, we have another type of mass term which comes from
the existence of the explicit chiral symmetry breaking due to
the current quark mass.
In the ordinary chiral perturbation theory the current quark mass
is included by the vacuum expectation value of the scalar external
source field ${\cal S}$.
For two flavor case this is given by
\begin{equation}
\langle {\cal S} \rangle =
{\cal M} =
\left( \begin{array}{cc}
m_u & \\
 & m_d
\end{array} \right)
\ .
\end{equation}
In the present analysis we work in the chiral limit, so that
we take
\begin{equation}
\langle {\cal S} \rangle = 0 \ .
\end{equation}
This scalar source field ${\cal S}$ combined with the
pseudoscalar source field ${\cal P}$ has the following
transformation property under the chiral symmetry:
\begin{equation}
{\cal S} + i {\cal P} \rightarrow
g_L \left( {\cal S} + i {\cal P} \right) g_R^\dag \ ,
\end{equation}
where $g_L$ and $g_R$ are elements of chiral
$\mbox{SU}(2)_{\rm L,R}$ groups.
To include ${\cal S}+i{\cal P}$ into the Lagrangian,
we define
\begin{equation}
\widehat{\cal M} \equiv \xi^\dag \,
\left( {\cal S} + i {\cal P} \right) \,\xi^\dag \ .
\end{equation}
This $\widehat{\cal M}$ transforms as
\begin{equation}
\widehat{\cal M} \rightarrow
  h(\pi,g_{\rm R},g_{\rm L}) \cdot \widehat{\cal M}
  \cdot
  h^\dag(\pi,g_{\rm R},g_{\rm L}) \ .
\end{equation}
Then, the term including this $\widehat{\cal M}$ field is written
as
\begin{equation}
{\cal L}_{\cal M} = \sum_{\vec{v}_F}
  C_{\cal M} \, \bar{\Psi}
  \left[ \widehat{\cal M} + \widehat{\cal M}^\dag \right]
\Psi \ ,
\label{Lag M}
\end{equation}
where $C_{\cal M}$ is a dimensionless real constant.~\footnote{
  We define the chemical potential $\mu$ at the chiral limit,
  so that we include the term in Eq.~(\ref{Lag M}) into the
  Lagrangian.
}
Note that, for constructing the above term,
we used the parity
invariance after summing over the Fermi velocity $\vec{v}_F$
(and integrating over $\vec{x}$ as usual):
The transformation properties of $\Psi$ and $\widehat{\cal S}$
under parity are
given by
\begin{eqnarray}
\Psi(x_0,\vec{x};\vec{v}_F)
\ \mathop{\rightarrow}_{P}\
\Psi(x_0,-\vec{x};-\vec{v}_F) \ ,
\nonumber\\
\widehat{\cal M}(x_0,\vec{x})
\ \mathop{\rightarrow}_{P}\
\widehat{\cal M}^\dag(x_0,-\vec{x})  \ .
\end{eqnarray}
Then,
\begin{eqnarray}
&&
\sum_{\vec{v}_F} \int d^4\vec{x}\,
\bar{\Psi}(x_0,\vec{x};\vec{v}_F)
\widehat{\cal M}(x_0,\vec{x})
\Psi(x_0,\vec{x};\vec{v}_F)
\nonumber\\
&& \quad
\ \mathop{\rightarrow}_{P}\
\sum_{\vec{v}_F} \int d^4\vec{x} \,
\bar{\Psi}(x_0,-\vec{x};-\vec{v}_F)
\widehat{\cal M}^\dag(x_0,-\vec{x})
\Psi(x_0,-\vec{x};-\vec{v}_F)
\nonumber\\
&& \quad
=
\sum_{\vec{v}_F} \int d^4\vec{x} \,
\bar{\Psi}(x_0,\vec{x};\vec{v}_F)
\widehat{\cal M}^\dag(x_0,\vec{x})
\Psi(x_0,\vec{x};\vec{v}_F)
\ .
\end{eqnarray}
Since the current quark masses are assigned as ${\cal O}(Q^2)$
in the ordinary chiral perturbation theory,
it is natural to assign ${\cal O}(Q^2)$
to the field $\widehat{\cal M}$.
Then, we count
the term in Eqs.~(\ref{Lag M})
as ${\cal O}(Q^2)$:
\begin{equation}
{\cal L}_{\cal M}
\ \sim \ {\cal O}(Q^2) \ ,
\end{equation}
or equivalently this term carries $\nu_i=3$ in the power counting
given in Eq.~(\ref{powernu}).

\subsection{Quark condensate}

The quark condensate can be read from the
one-point function of the external source field ${\cal S}$.
In the present paper we decompose ${\cal S}$ as
\begin{equation}
{\cal S} = \frac{1}{N_f} {\cal S}^0 +
   \sum_{a=1}^{N_f^2-1} {\cal S}^a T_a \ ,
\end{equation}
where $T_a$ is the generator of $\mbox{SU}(N_f)$
normalized as $\mbox{tr}[ T_a T_b ]= \frac{1}{2} \delta_{ab}$.
Then, the one-point function of ${\cal S}^0$ provides the
quark condensate per one flavor.
{}From the interaction in Eq.~(\ref{Lag M}), one-loop correction
to one-point function of ${\cal S}^0$ is expressed as
\begin{equation}
\Delta = - 4 C_{\cal M}
\sum_{\vec{v}_F} \int \frac{d^4 l}{ i (2\pi)^4}
  \frac{1}{ - V_F\cdot l - i \epsilon l_0 }
\ .
\label{def Delta}
\end{equation}
This integral is divergent, so that we here define the
regularization to perform the integral.
We first replace the momentum integral as
in Eq.~(\ref{rep mom int}).
Then, we introduce the cutoff for $l_{\parallel}$ and
$l_0$ as
$- \Lambda_{\parallel} < l_{\parallel} < \Lambda_{\parallel}$
and
$- \Lambda_0 < l_0 < \Lambda_0$.
By using this regularization method, Eq.~(\ref{def Delta}) is
written as
\begin{eqnarray}
\Delta
&=&
 - 4 C_{\cal M}
\sum_{\vec{v}_F}
\int \frac{d^2\vec{l}_{\perp}}{(2\pi)^2}
\int \frac{ d l_{\parallel}}{2\pi}
  \theta( \Lambda_{\parallel} - \vert l_{\parallel} \vert )
\int \frac{ d l_0 }{2\pi i}
  \theta( \Lambda_0 - \vert l_0 \vert )
  \frac{1}{ - l_0 + \bar{v}_F l_{\parallel}- i \epsilon l_0 }
\ .
\label{reg Delta}
\end{eqnarray}
Performing the $l_0$ integral, we obtain
\begin{eqnarray}
\Delta
&=&
 - 4 C_{\cal M}
\sum_{\vec{v}_F}
\int \frac{d^2\vec{l}_{\perp}}{(2\pi)^2}
\int \frac{ d l_{\parallel}}{2\pi}
  \theta( \Lambda_{\parallel} - \vert l_{\parallel} \vert )
\nonumber\\
&& \
{}\times
\Biggl[
  - \theta( - \bar{v}_F l_{\parallel} )
  \theta( \Lambda_0 - \vert \bar{v}_F l_{\parallel} \vert )
  + \frac{1}{2}
\nonumber\\
&& \qquad
  {}+ \frac{1}{2\pi i}
  \left\{
    \ln \left(
      \Lambda_0 + ( \bar{v}_F l_{\parallel} )(1-i\epsilon)
    \right)
    -
    \ln \left(
      \Lambda_0 - ( \bar{v}_F l_{\parallel} )(1-i\epsilon)
    \right)
  \right\}
\Biggr]
\ .
\end{eqnarray}
By taking the $\Lambda_0 \rightarrow \infty$ limit,
this expression is reduced to the following finite form:
\begin{eqnarray}
\Delta
&=&
 - 4 C_{\cal M}
\sum_{\vec{v}_F}
\int \frac{d^2\vec{l}_{\perp}}{(2\pi)^2}
\int \frac{ d l_{\parallel}}{2\pi}
  \theta( \Lambda_{\parallel} - \vert l_{\parallel} \vert )
\left[
  - \theta( - \bar{v}_F l_{\parallel} )
  + \frac{1}{2}
\right]
\ .
\end{eqnarray}
Now, performing the $l_{\parallel}$ integral, we obtain
\begin{equation}
\Delta = 0 \ .
\end{equation}
This implies that the number density of the fluctuation mode
at one-loop level is zero in chiral limit.
It is a kind of a symptom that we choose a proper vacuum in matter.

\section{Discussions}
\label{disc}

As the density of the nuclear medium increases,
the Fermi momentum $p_F$ emerges
as an additional scale of the system, which
deserves to be treated differently from the small scale $Q$ of HBChPT,
for a more transparent and better chiral expansion scheme.
In this work, we have considered the cases where $p_F$ is much larger than $Q$
but still much smaller than the chiral symmetry breaking scale,
$Q\ll p_F \ll \Lambda_\chi$.

We studied
the vector and the axial vector correlators in the nuclear medium,
and we could calculate the Debye screening mass, pion velocity and the
modification of the pion decay constant.
The density-dependence of those quantities is found to be highly
different from that of the HBChPT, even at the moderate density
region.
Instead of having the leading linear density dependence as was the case in HBChPT,
we have observed rather mild dependence.
This may be understood by a following simple consideration:
At very low density, the
response of the external probe would be proportional to the number
of particles in the system, or to the volume of the Fermi sea.
When density increases, however, excitations are limited around 
the Fermi surface
due to the Pauli blocking,
and the response would now be proportional to the area of the Fermi surface.

The present study enables us to describe the nuclear matter
up to the density comparable to $\rho_0$, the normal nuclear matter density.

To extend the applicable region of the theory,
we need to do RPA type of calculations to sum all the
contributions with respect to $p_F/\Lambda_\chi$,
for a given order of $Q/\Lambda_\chi$.
This step will result in a very rapid convergence for the
processes where $Q/\Lambda_\chi\ll 1$, regardless of the density.
The actual calculation can in fact be done rather easily.
But this step makes our results to be dependent on the coefficients
of the four-fermion interactions.
And thus we need a more complete study including the renormalization
of those coefficients to have any meaningful predictions.
In addition to this, as the density increases, the intrinsic density dependence
of our theory
can play a crucial role, and thus should be properly treated.
The notion of the {\it decimation}~\cite{rho-brown},
and the renormalization-group running and additional matching conditions
might be necessary
to address this issue.
Studies on these subjects are left for the future work.

In the high density region in which the RPA is needed, the
perturbative structure of the chiral effective theory will be most
likely modified with the necessity of choice of the relevant degrees
of freedom that we play with. Indeed, we may need to introduce the
light vector meson degrees of freedom as claimed to do for the
matching with the QCD in the phase of the Vector
Manifestation~\cite{HY:VM}. Even with the chiral partner of the pion
is incorporated, the perturbative scheme should stay unchanged as
shown in the counting rule.

However, when the chiral phase transition is approached, we have to
reconsider the change of the representative scales of the medium from
what is considered in this work. Namely, the $4\pi f_\pi$ is
reduced eventually to be zero,
and thus it cannot serve as the ``heavy scale'' of the system.
In this case, the relevant expansion parameter would be
the ratio of the typical momentum transfer compared to the chemical potential,
$Q/\mu$, which results in a theory with the same scale dependence as
the HDET~\cite{schaefer}.  The transmutation of the scales could be
realized by considering the second decimation as is suggested by Brown
{\it et al}~\cite{BLR:decimation}.

\section*{Acknowledgment}

The authors would like to thank Mannque Rho and
Deog Ki Hong for useful discussions and comments.
Four of the authors (M.Harada, D.P.Min, T.S.Park and C.Sasaki) would like to
thank the members of ECT* for warm hospitality during their stay,
where the final part of this work was done.
M.Harada and C.Sasaki would like to thank the members of
nuclear theory group in Seoul National University
for warm hospitality. The work of M.Harada, D.-P.Min, C.Song was supported in part by the BK21 project of the MOE, Korea and by BP Korea program, KOFST. 
The work of M.Harada and C.Sasaki is supported in part by
the JSPS Grant-in-Aid for Scientific Research (c) (2) 16540241,
and by
the 21st Century COE
Program of Nagoya University provided by Japan Society for the
Promotion of Science (15COEG01). The work of DPM is supported by KOSEF-R01-1999-000-00017-D, KRF-2001-015-DP0085.

\end{document}